\UseRawInputEncoding
\documentclass[fleqn,usenatbib]{mnras}

\usepackage[T1]{fontenc}
\usepackage{ae,aecompl}
\usepackage{newtxtext,newtxmath}
\usepackage{xcolor}
\usepackage{graphicx}	% Including figure files	
\usepackage{subfig}
\usepackage{epsfig}
\usepackage{wrapfig}
\usepackage{times,longtable,color}
\usepackage{siunitx}
\usepackage{amsmath}
\newcommand {\apgt} {\ {\raise-.5ex\hbox{$\buildrel>\over\sim$}}\ }
\newcommand {\aplt} {\ {\raise-.5ex\hbox{$\buildrel<\over\sim$}}\ }

\title[NuSTAR reveals the hidden nature of SS433]
{NuSTAR reveals the hidden nature of SS433}

\author[M. Middleton et al.]
{M. J. Middleton,$^{1}$, D. J. Walton $^{2}$, W. Alston$^{2,3}$, T. Dauser$^{4}$, S. Eikenberry$^{5}$, Y-F Jiang$^{6}$,\newauthor A. C. Fabian$^{2}$, F. Fuerst$^{7}$, M. Brightman$^{8}$, H. Marshall$^{9}$, M. Parker$^{3}$, C. Pinto$^{2, 10}$,\newauthor F. A. Harrison$^{7}$, M. Bachetti$^{10}$, D. Altamirano$^{1}$, A. J. Bird$^{1}$, G. Perez$^{5}$, J. Miller-Jones$^{11}$,\newauthor P. Charles$^{1}$, S. Boggs$^{12}$, F. Christensen$^{13}$, W. Craig$^{7}$, K. Forster$^{7}$, B. Grefenstette$^{7}$,\newauthor C. Hailey$^{14}$, K. Madsen$^{7}$, D. Stern$^{15}$, W. Zhang$^{16}$\\
\\
1. Department of Physics and Astronomy, University of Southampton, Highfield, Southampton SO17 1BJ, UK\\
2. Institute of Astronomy, University of Cambridge, Madingley Road, Cambridge CB3 0HA, UK\\
3. European Space Agency (ESA), European Space Astronomy Centre (ESAC), E-28691 Villanueva de la Ca\~{n}ada, Madrid, Spain\\
4. Remeis Observatory \& ECAP, Universität Erlangen-Nürnberg, Sternwartstr. 7, 96049 Bamberg, Germany\\
5. Department of Astronomy, University of Florida, Gainesville, FL 32611, USA\\
6. Center for Computational Astrophysics, Flatiron Institute, NY 10010, USA\\
7. Quasar Science Resources S.L for European Space Agency (ESA), European Space Astronomy Centre (ESAC), Camino Bajo del Castillo s/n, 28692 Villanueva de la Cañada, Madrid, Spain\\
8. Cahill Center for Astrophysics, California Institute of Technology, 1216 East California Boulevard, Pasadena, CA 91125, USA\\
9. Kavli Institute for Astrophysics and Space Research, Massachusetts Institute of Technology, 77 Massachusetts Ave., Cambridge, MA 02139, USA\\
10. INAF-Osservatorio Astronomico di Cagliari, via della Scienza 5, I-09047 Selargius, Italy\\
11. International Centre for Radio Astronomy Research, Curtin University, GPO Box U1987, Perth, WA 6845, Australia\\
12. Space Sciences Laboratory, University of California, Berkeley, CA 94720-7450, USA\\
13. DTU Space-National Space Institute, Technical University of Denmark, Elektrovej 327, DK-2800 Lyngby, Denmark\\
14. Columbia Astrophysics Laboratory, Columbia University, New York, NY 10027, USA\\
15. Jet Propulsion Laboratory, California Institute of Technology, 4800 Oak Grove Drive, Mail Stop 169-221, Pasadena, CA 91109, USA\\
16. X-ray Astrophysics Laboratory, NASA Goddard Space Flight Center, Greenbelt, MD 20771, USA
}

\pagerange{\pageref{firstpage}--\pageref{lastpage}} \pubyear{2021}
\long\def\symbolfootnote[#1]#2{\begingroup\def\thefootnote{\fnsymbol{footnote}}\footnote[#1]{#2}\endgroup} 
\def\ga{\mathrel{\hbox{\rlap{\hbox{\lower4pt\hbox{$\sim$}}}{\raise2pt\hbox{$>$}}
}}}

\begin{document}
\topmargin = -0.5cm

\maketitle

\label{firstpage}

\begin{abstract}
SS433 is the only Galactic binary system known to persistently accrete at highly super-critical (or hyper-critical) rates, similar to those in tidal disruption events, and likely needed to explain the rapid growth of those very high redshift quasars containing massive SMBHs. Probing the inner regions of SS433 in the X-rays is crucial to understanding this system, and super-critical accretion in general, but is highly challenging due to obscuration by the surrounding wind, driven from the accretion flow. {\it NuSTAR} observed SS433 in the hard X-ray band across multiple phases of its 162 day super-orbital precession period. Spectral-timing tools allow us to infer that the hard X-ray emission from the inner regions is likely being scattered towards us by the walls of the wind-cone. By comparing to numerical models, we determine an intrinsic X-ray luminosity of $\ge$ 2$\times$10$^{37}$ erg/s and that, if viewed face on, we would infer an apparent luminosity of  $>$ 1$\times$10$^{39}$ erg/s, confirming SS433's long-suspected nature as an ultraluminous X-ray source (ULX). We present the discovery of a narrow, $\sim 100$~s lag due to atomic processes occurring in outflowing material travelling at least 0.14-0.29c, which matches absorption lines seen in ULXs and -- in the future -- will allow us to map a super-critical outflow for the first time.
\end{abstract}


\section{Introduction}

Discovered in 1977 from its bright H$_{\alpha}$ emission (Stephenson \& Sanduleak 1977), SS433's defining characteristics are undoubtedly the helical motion of highly-collimated jets of plasma launched from its innermost regions, and mass-loaded, non-polar outflows (Fabian \& Rees 1979; Margon et al. 1979) which together inflate the surrounding W50 supernova remnant. Knots in SS433's jet can be resolved at radio frequencies using very long baseline interferometry (VLBI) and indicate the presence of highly relativistic electrons (Vermeulen et al. 1987), while the baryon content is revealed by emission lines ranging from H and He lines in the optical through to highly ionised Fe lines in the X-rays (Kotani et al. 1994; Marshall et al. 2013). The Doppler shifts of the lines indicate precession of the accreting system with a period of $\approx$ 162 days, also seen in optical (HeII) emission lines originating from the non-polar wind (Fabrika 1997). Both the jets and winds carry a large kinetic luminosity ($>$ 10$^{38}$ erg/s, e.g. Marshall et al. 2002), which requires extraction of energy via accretion onto a compact object. While the nature of the compact object in SS433 remains somewhat unknown (although dynamical arguments suggest the presence of a black hole - Blundell, Bowler \& Schmidtobreick 2008), the rate of mass transfer from the companion star, as inferred from the IR excess (Shklovskii 1981; Fuchs et al. 2006), is thought to be $\sim$1$\times$10$^{-4}$ M$_{\odot}$/year, orders of magnitude in excess of the Eddington limit for any plausible stellar remnant ($>$ 300 times the Eddington mass accretion rate for a typical stellar mass black hole of around 10 M$_{\odot}$). Classical theory and radiation magneto-hydrodynamic (RMHD) simulations agree that such `super-critical' rates of accretion will lead to a radiatively supported, large scale height (H/R $\approx$ 1, where H is the height of the disc at distance R from the compact object) accretion disc with powerful winds launched from the surface at mildly relativistic speeds (Shakura \& Sunyaev 1973; Poutanen et al. 2007; Ohsuga \& Mineshige. 2011; Takeuchi et al. 2013; Jiang et al. 2014; Sadowski et al. 2014). 

Super-critical systems such as ultraluminous X-ray sources (ULXs: Kaaret, Feng \& Roberts 2017) and tidal disruption events (TDEs, e.g Rees 1988; van Velzen \& Farrar 2014) in their early phases, are predicted to be extremely bright at a range of wavelengths due to the intrinsic luminosity following an L$_{\rm Edd}[1+{\rm ln}(\dot{m}_{0})]$ dependence (where $\dot{m}_{0}$ is the accretion rate in Eddington units, see Shakura \& Sunyaev 1973). This dependence alone can allow the Eddington luminosity to be exceeded by a factor of a few (ignoring the energy spent in launching outflows) and indeed, SS433 is known to have optical (and UV) luminosities in excess of 10$^{40}$ erg/s (Dolan et al. 1997; Waisberg et al. 2019). At higher energies, these systems are expected to be exceedingly X-ray bright due to collimation (`geometrical beaming' - see King et al. 2001; 2009) by the super-critical disc and optically thick wind. In this regard SS433 is remarkably X-ray {\it faint} at only 10$^{35} - 10^{36}$ erg/s (at an estimated distance of $\approx$ 6~kpc: Lockman, Blundell \& Goss 2007), several orders of magnitude below the Eddington luminosity for a stellar remnant. This X-ray faintness implies that at no point during the system precession do we see down to the central engine, consistent with a large scale-height inflow/wind and a mean inclination of 78.8 degrees to our line-of-sight (Margon \& Anderson 1989). Given the predictions that SS433 is likely to be an edge-on ULX (fitting neatly into the picture where ULXs exist as a continuum of inclined super-critical sources, e.g. Poutanen et al. 2007; Middleton et al. 2015), confirming its intrinsic ULX-like properties and obtaining a reliable estimate for the intrinsic X-ray luminosity is important but has not been straightforward (though notable attempts have been made, e.g. from studying the effects of irradiation on aligned molecular clouds: Khabibullin \& Sazonov 2016). 

Below 20 keV, SS433's X-ray spectrum is well known to be dominated by thermal Bremsstrahlung emission from the baryon-loaded jets (e.g. Brinkmann, Kotani \& Kawai 2005), and reveals the nature of the plasma from which the emission lines originate (Marshall et al. 2013). Above 20 keV there is a hard excess of emission, previously  detected by both ESA's {\it INTEGRAL} (Cherepashchuk et al. 2013) and JAXA/NASA's {\it Suzaku} satellite (Kubota  et al. 2010). This hard X-ray emission may naturally arise from some hotter component of the jet (e.g. Medvedev, Khabibullin \& Sazonov 2020), Compton scattering of a seed photon field off electrons in the jet (Fabrika 2004), Comptonisation of some photon field by a plasma of hot electrons in the wind cone (e.g. Cherepashchuk et al. 2005, 2007, 2013; Krivosheyev et al. 2009) and/or reflection of the radiation emitted from within the wind-cone (e.g. Medvedev \& Fabrika 2010). Should emission from the jet dominate, we would predict a single continuous spectral component with a range of temperatures, reflecting the increasing temperature of the jet plasma approaching the acceleration point (Marshall, Canizares \& Schulz 2002), while Comptonisation or reflection would appear as a spectral component distinct from that of the jet. In both situations we might expect absorption by material associated with the outflowing wind along our line-of-sight (as has been detected in the case of both ULXs and TDEs: Middleton et al. 2014; Walton et al. 2016; Pinto et al. 2016, 2017; Kosec et al. 2018; Kara et al. 2018). 

Reliably determining the nature of the hard emission in SS433 is clearly important for understanding the system, but time-averaged approaches often result in degenerate solutions without prior assumptions for the underlying physics. Applying time-resolved methods can allow these degeneracies to be circumvented (see Uttley et al. 2014). Here we present the results of our {\it NuSTAR} campaign which provides new insights and allows constraints on the intrinsic luminosity to be obtained. 

\begin{table*}
\begin{center}
\begin{minipage}{120mm}
\bigskip
\caption{Observational details for the {\it NuSTAR} campaign.}
\begin{tabular}{l|c|c|c|c|c|c|c}

\hline 
\hline
Obs    &  OBSID    & JD   & T$_{\rm exp}$ (ks)   & $\Phi_{\rm prec}$ & $\Phi_{\rm orb}$ & $\Sigma$ (deg) & L$_{\rm ref}~~(10^{35}$  erg s$^{-1}$)  \\
\hline
1          &  30002041004  & 2456961  & 25  & 0.85 & 0.28 & 68 & 5.20 $\pm 0.05$  \\
2         &  30002041006  & 2456974  & 29  & 0.93 & 0.28 & 60 & 12.05  $\pm 0.06$ \\
3         &  30002041008  & 2456987  & 28  & 0.01 & 0.28 & 59 & 14.56 $_{-0.09}^{+0.05}$ \\
4          &  30002041012  & 2457078  & 21  & 0.58 & 0.27 & 96 & 2.68 $\pm 0.04$ \\ 
5          &  30002041014   & 2457093  & 26  & 0.66 & 0.35 & 90 & 2.06 $_{-0.05}^{+0.02}$  \\
6          &  30002041016  &  2457105 &  30 & 0.74 & 0.32 & 81 & 3.19 $_{-0.05}^{+0.02}$  \\
7          &  30002041018  & 2457131  & 27  & 0.90 & 0.31 & 63 & 12.30 $\pm 0.06$ \\
8          &  30002041020  &  2457209 & 27  & 0.38 & 0.21 & 94 & 2.95 $_{-0.05}^{+0.04}$ \\
\hline
\end{tabular}

Notes: Observational details of the {\it NuSTAR} SS433 campaign. Across the Table we report the Observation IDs (OBSIDs), date of the observation (in Julian days), exposure time (noting that the observation length is twice this due to {\it NuSTAR}'s low Earth orbit), precessional and orbital phases ($\Phi_{\rm prec}$ and $\Phi_{\rm orb}$, from established ephemerides - Eikenberry et al. 2001), the inferred inclination of the wind-cone from the kinematic model ($\Sigma$) and the observed luminosity in the reflected component ($L_{\rm ref}$) assuming a distance to the source of $\approx$ 6 kpc (Lockman et al. 2007). The latter are plotted versus inclination in Figure 10.
\end{minipage} 
\end{center}
\end{table*}

\section{The SS433 campaign}

NASA's {\it NuSTAR} X-ray satellite (Harrison et al. 2013) observed SS433 in the 3-79 keV band over several precessional phases, with observations scheduled to avoid the eclipse by the companion star. The observing dates and spectroscopic precessional phases - based on the well-established optical ephemerides (Eikenberry et al 2001) - are provided in Table 1. All observations obtained by {\it NuSTAR} were visually inspected for stray light with no contamination identified. We therefore extracted spectral and timing products using the {\it NuSTAR} pipeline, {\sc nupipeline} v1.9.2 with 100 arcsecond source and background extraction regions and default SAA exclusion options. We further re-binned the spectral data to oversample the response by a factor 3 and such that each energy bin has a signal-to-noise ratio of 6. This resulted in sufficient counts/energy bin for chi-squared fitting of the energy spectrum. Two additional observations to those reported in Table 1 were taken by {\it NuSTAR}, however, one of these only had data in a non-standard mode and was excluded while one showed unusual behaviour which will be discussed in a future work.

The raw spectra, unfolded through a power-law of zero index and unity normalisation (to avoid bias) are shown in Figure 1. Even though the source is relatively X-ray faint, the remarkably high-quality spectra (a result of the high sensitivity of {\it NuSTAR}) allow the changing emission properties of the system on long timescales to be studied in detail, and clearly show the characteristic Fe XXV/XXVI emission lines moving with phase, and the hard excess becoming increasingly peaked above 20~keV with increasing flux. The source's emission is well known to vary on shorter, intra-observational ($<$ 25 ks) timescales (Revnivtsev et al. 2004, 2006; Atapin et al. 2015) which has allowed us to obtain insights and inform our decisions when it comes to modelling the time-averaged spectra.

\begin{figure}
\includegraphics[trim=110 150 60 80, clip, width=8.5cm]{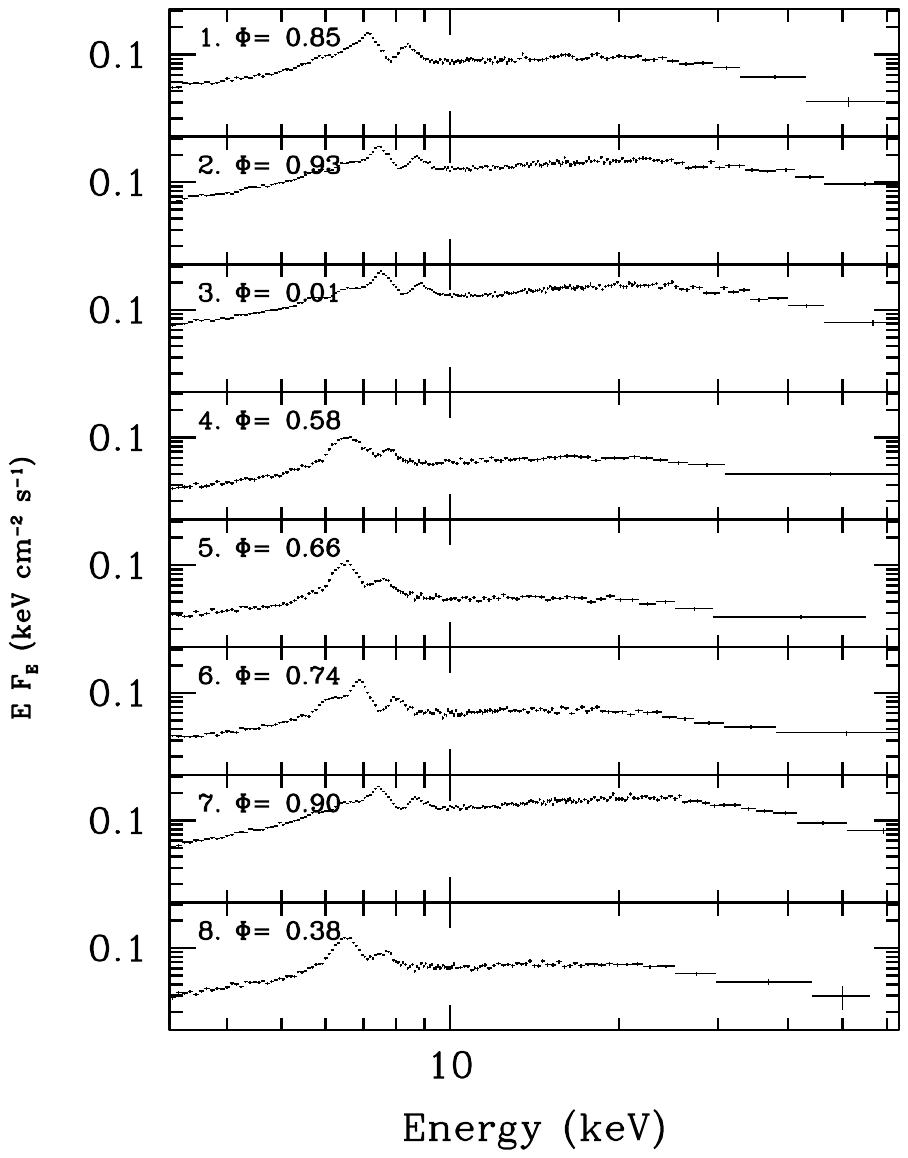}
    \caption{The eight {\it NuSTAR} (FPMA only) spectra of SS433 considered here, unfolded through a power-law of zero index and unity normalisation.}
    \label{fig:flowchart}
\end{figure}

\subsection{The cross-spectrum}

Before studying the time-averaged spectrum, we extract and explore the cross-spectrum. Such analyses when applied to X-ray binaries and AGN (see Uttley et al. 2014) can lead to profound and otherwise unavailable insights. 

From two evenly-sampled time series x(t), y(t) we can compute the cross-spectrum in frequency (f) space, C$_{\rm xy}$(f) = X(f)*Y(f) = $|X||Y|$e$^{i(\phi_{\rm y} - \phi_{\rm x})}$) where * denotes the complex conjugate of the Fourier transforms X(f), Y(f), with amplitude and phase ($\phi$). We estimated cross-spectral products by first averaging the complex C$_{\rm xy}$(f) values over $m$ non-overlapping segments of a given observation's time-series, and then averaging in geometrically spaced frequency bins (see Uttley et al. 2014 for more details on the use of the cross-spectrum).

\begin{figure*}
\includegraphics[trim=50 120 0 100, clip, width=16cm]{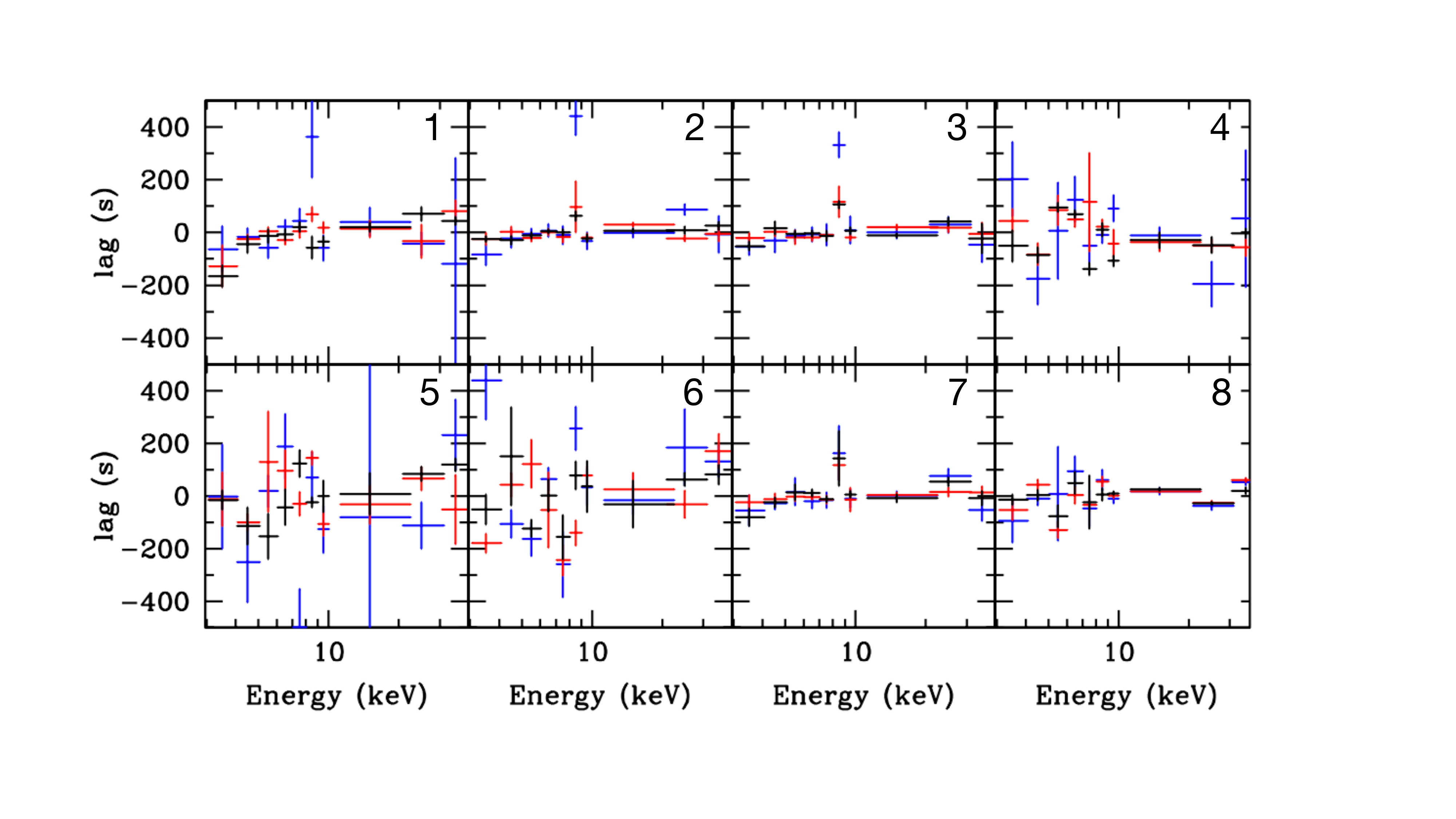}
    \caption{Energy-lag spectra for each observation of SS433 as a function of Fourier frequency, where blue = 0.5-1.5 mHz, red = 1.5-2.5 mHz and black = 2.5 - 3.5 mHz. From top left to top right: Obs 1-4, and from bottom left to bottom right: Obs 5-8. There is clearly a significant lag in four observations (1, 2, 3 and 7) in the 8-9 keV bin, indicating an atomic process.}
    \label{fig:flowchart}
\end{figure*}

\subsection{Time lags - imprint of a wind}

From the argument of C$_{\rm xy}$(f) we obtain a phase shift (lag) of $\Phi (f) = {\rm arg} \langle C_{\rm xy}(f)\rangle$ which can be transformed into the corresponding time lag: $\tau(f) = \Phi(f) /2\pi f$.  This gives the (time-averaged) frequency-dependent time lag between any correlated variations in x(t) and y(t). Errors on $\tau(f)$ are estimated using standard formulae (Vaughan \& Nowak 1997; Uttley et al. 2014). For a given Fourier frequency we can therefore compute the time lag between a comparison energy band y(t) and a broad reference band x(t). The reference band is typically chosen to have a high signal-to-noise ratio (S/N) and high variability power so time delays between any correlated variations in bands with weaker S/N can be recovered; a fairly broad reference band is therefore typically utilized. Where they overlap, the band of interest is removed when computing the lag between the two, in order to avoid any correlated noise (Uttley et al. 2014). A more positive lag value indicates the comparison-band lags the reference band.

\begin{figure*}
\includegraphics[trim=50 100 0 100, clip, width=20cm]{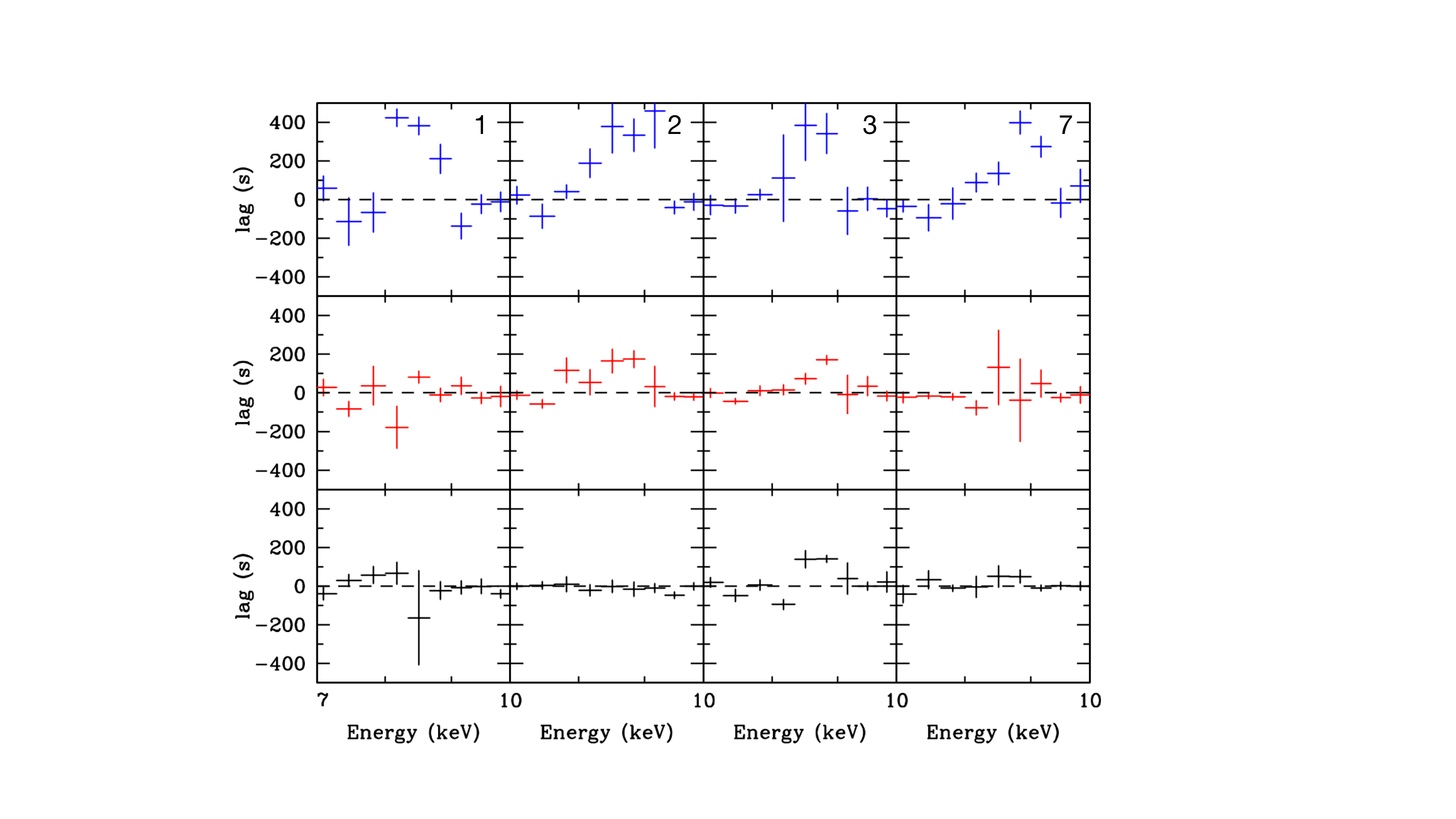}
    \caption{Higher resolution energy-lag spectra (showing a zoom-in on the 7-10 keV band)
    for observations 1, 2, 3 and 7 as a function of Fourier frequency, where blue = 0.5-1.5 mHz, red = 1.5-2.5 mHz and black 2.5 - 3.5 mHz. The lags are clearly peaked at $\approx$ 400s and change with precessional phase and with a clear frequency dependence.}
    \label{fig:flowchart}
\end{figure*}

We use a segment length of 1 ks so we have m $\ge$ 25 segments per observation, giving enough estimates for the errors to be Gaussian distributed. We use the 3-6 keV band as reference and summed FPMA and FPMB lightcurves. The power spectral density has significant power above the noise at low frequencies (e.g. Atapin et al. 2015), such that we can explore the lags in the lowest three frequency bands: 0.5 - 1.5, 1.5 - 2.5 and 2.5 - 3.5 mHz. The lag versus energy spectra are shown in Figure 2; there is clearly a strong signature of a positive lag at most (if not all) frequencies in observations 1, 2, 3 and 7 in the 8-9 keV bin while the other observations show a noisier, more complicated behaviour. This could be due to changes in the power as a function of precessional phase (see Atapin et al. 2015) or the physical mechanism by which the lag is imprinted. 

In Figure 3, we show a zoom-in of the 7 - 10 keV portion of the lag-energy spectrum for observations 1, 2, 3 and 7. The shape of the lag appears to change between observations, appearing peaked with a total width of $\approx$ 1 keV, with the maximum lag diminishing from a maximum of $\approx 400~s$ with increasing Fourier frequency.

\begin{figure}
\includegraphics[trim=0 0 0 0, clip, width=8cm]{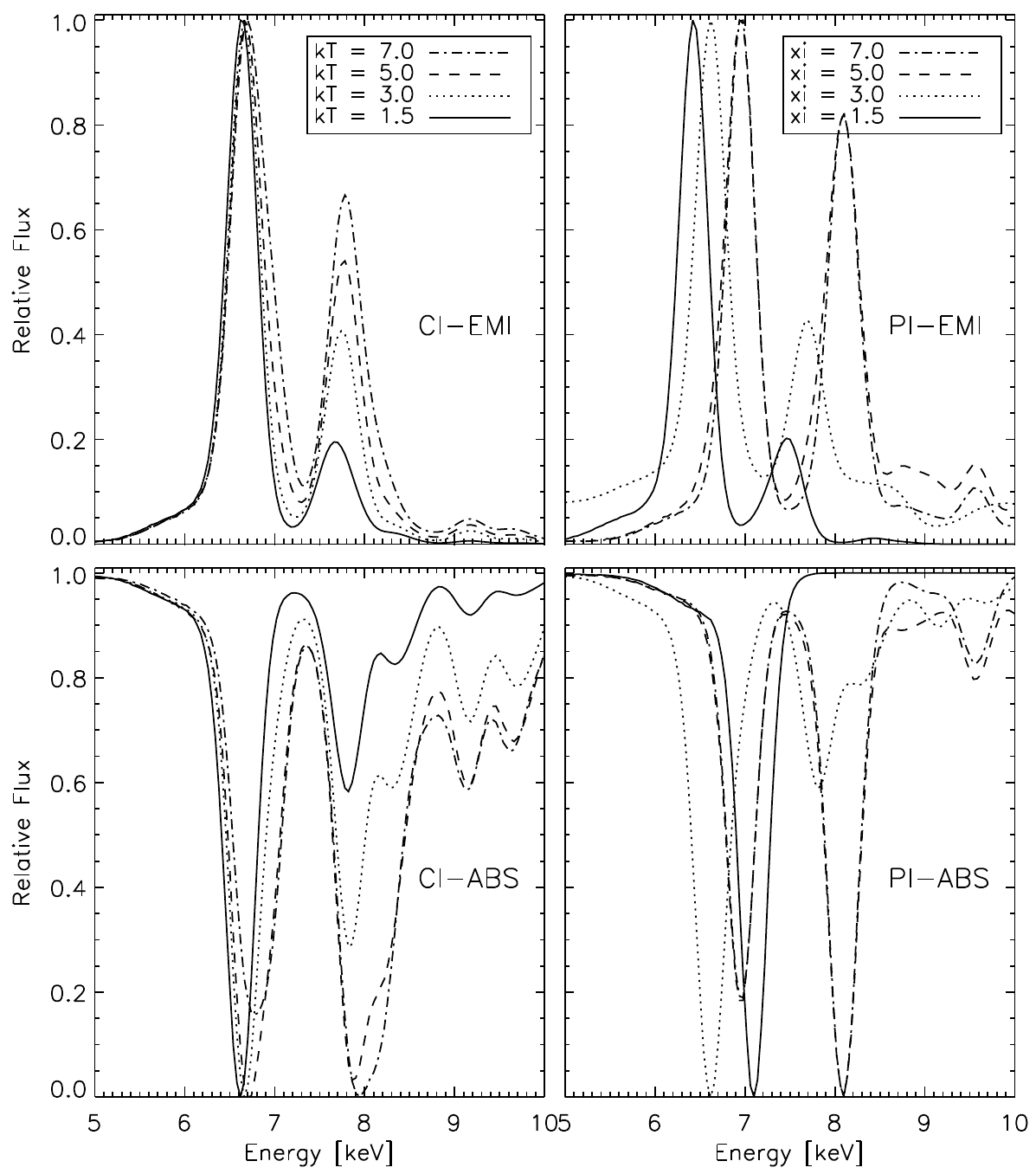}
    \caption{Using an assumed SED (from Pinto et al. 2020), gas is photo-ionised (PI) or assumed to be in collisionally ionised (CI) equilibrium, and the ionisation parameter or temperature are varied to explore the impact on absorption(ABS)/emission(EMI) line emissivities assuming the super-solar abundances in SS433's jet.}
    \label{fig:flowchart}
\end{figure}

A lag over such a narrow energy range is only likely to result from emission/absorption processes in atomic transitions. There are two primary possibilities for creating line emission, collisionally excited emission/absorption as seen in the jet, and photo-absorption and emission due to recombination/relaxation. Notably, the lag appears to be well isolated in the 8-9~keV bin and does not appear to match the number or energy range of the Fe XXV/XXVI emission lines associated with the jet (specifically at energies below 8 keV - see Figures 1, 5 \& 6). 

\subsection{Potential origins for the lag}

{\it Photo-electric edge in a jet sheath}: Kubota et al. (2007), identified the possible presence of a blue-shifted Fe K photoelectric edge in the {\it XMM-Newton} spectra of SS433, suggesting this could be associated with absorption in a thin co-moving sheath around the jet. However, there are potential obstacles to associating the lags we have discovered with such an edge. In a photo-electric edge, photons are absorbed across a range of energies and are effectively destroyed. As a result, it is only possible to create a lag due to propagation if a velocity gradient is present across the absorbing material such that the edge energy moves to lower energies with distance (i.e. a slower moving sheath around a faster moving one). As a consequence, the fluorescence line from faster moving material can sit within the photoelectric edge of the slower moving material and the emitted photon can then be absorbed. The result is that the largest lag should correspond to the slowest moving material. Given the rest-frame energy of the Fe K edge (at maximum optical depth) is 7.11 keV, then the lags from 8 - 9.3 keV would lead us to infer line-of-sight velocities of 0.12 to 0.26c, the upper end of which is considerably higher than the projected velocity of the jet at these inclinations. Regardless of the implied line-of-sight velocity, the magnitude of the lag of $\sim$ 100s would demand a thickness of $D \sim c\times lag/\tau$ where $\tau$ is the optical depth of the sheath plasma. Assuming $\tau \sim 1$ then implies $D \sim 10^{13}$ cm which is the suggested length of the X-ray jet. Given the narrow opening angle of the jet ($<$ a few degrees, see Begelman et al. 1980; Vermeulen et al. 1993; Marshall, Canizares \& Schulz 2002), even at the lowest inclination angle of $\approx 60$ degrees, it seems unlikely that such distances would be available.

{\it Recombination in a jet sheath}: It is also possible to imprint a lag through a response to a change in the irradiating flux (i.e. ionisation and recombination timescales: see Silva, Uttley \& Constantini 2016). Recombination would lead us to expect lags around $t_{\rm rec} \sim 0.3(T_{5}^{1/2}/n_{14}Z^{2})$~s (where $T_{5}$ is the temperature of the absorbing plasma in units of $10^{5}$K, $n_{14}$ is the ion number density in units of $10^{14} $cm$^{-3}$ and $Z$ is the atomic number of the absorbing species, see Jimenez-Garate et al. 2002). Should the material in the sheath be of similar density to that of the jet, then the number density is estimated to be $n = 10^{13} - 10^{14} $cm$^{-3}$ (Marshall et al. 2002). Even for a temperature as high as that of the jet ($\sim 10 $keV), we would expect recombination timescales of less than a second. 

{\it Resonant lines}: A more natural explanation for the lag may be the propagation of photons through a less collimated outflowing medium due to resonant absorption and emission (e.g. Middleton et al. 2019) which must take place in partially ionised winds. As the nature of the wind along our line-of-sight and the amount of variability we detect are inclination dependent (e.g. Atapin et al. 2015), this may then contribute to the strength and clarity of the feature as a function of precessional phase (Table 1). The likely association would be with blue-shifted Fe XXV/XXVI resonance lines for which the rest-frame energies are 6.7 and 6.97 keV respectively (Verner, Verner \& Ferland 1996), implying an outflow velocity in the line-of-sight ranging from 0.14 - 0.29c (lag at 8 keV), to 0.28 - 0.32c (lag at 9.3 keV), consistent with absorption lines detected in ULXs (Pinto et al. 2016). Given that the jet velocity in the line-of-sight is known to never reach such values (Eikenberry et al. 2001), this again supports the origin in absorbing material which is embedded in a less collimated (but similarly fast) outflow, i.e. a wind.

To explore the conditions for producing a single resonant line (or blend across a 1 keV bin) we use SPEX v 3.05 to simulate the effect of an irradiating SED which is hidden from view. As we do not have strong constraints on the intrinsic SED of SS433, we opt for that of the closer-to-face-on ULX, NGC 1313 X-1 (see Pinto et al. 2020, Walton et al. 2020) which is known to produce powerful outflows (Middleton et al. 2015; Pinto et al. 2016; Walton et al. 2016). We adopted the standard Lodders et al. (2009) abundance scale with elemental abundances of 2 $\times$ solar for Fe, and 20 $\times$ solar for Ni respectively (see section 2.4). We created the model line spectra in Figure 4 for collisionally excited absorption/emission across a range of temperatures, and photo-ionised absorption/emission for a range of ionisation parameters (note we do not include any velocities as we are interested only in the conditions under which single lines appear) which have been matched to the resolution of the {\it NuSTAR} detectors. It is apparent that, in order to yield only one line, the ionisation parameter would have to be low (log$\xi \lesssim$ 2-3 erg cm s$^{-1}$) or the gas would need to be low temperature (kT $<$ 1.5 keV). Given the expected irradiation (and high ionisation - see Medvedev \& Fabrika 2010) and the high temperature of the surrounding gas, a more likely explanation may be that the outflow has a far lower relative abundance in Ni compared to the jets. This has been suggested by previous studies (Medvedev et al. 2018) which invoke the presence of a neutron star (see also Goranskij 2011) on which the Ni is produced.

{\it Recombination in the wind}: An alternative mechanism for creating the lag could be recombination of the partially ionised plasma in the wind (rather than jet sheath), responding to changes in the irradiating continuum (Silva et al. 2016). As the density may be substantially lower than that in the jet ($\approx 4 \times 10^{6} $cm$^{-3}$: Fabrika 2004), a recombination timescale of hundreds of seconds is possible under the right conditions. However, the reason for the precessional phase dependence is then less clear (recombination should lead to a lag in the isotropic line emission, requiring those regions to somehow become obscured).

In light of the above possibilities and constraints, regardless of the exact nature of the lag, it would seem that an origin in the wind is the more natural explanation for the lags we have discovered in SS433.

\subsection{The covariance spectrum}

We can use the cross-spectrum to investigate the energy dependence of the variability (rather than only the lag between bands) across a range of Fourier frequencies. The covariance spectrum is analogous to the rms-spectrum (Revnivtsev, Gilfanov \& Churazov 1999, Wilkinson \& Uttley 2009) where the amplitude of $C_{\rm xy}$ (f) (normalised by $|X(f)|^{2}|Y(f)|^{2}$) gives the coherence - the degree of linear correlation between x(t) and y(t) as a function of Fourier frequency - which takes values from 0 to 1. We can use the coherence to obtain a Fourier frequency resolved covariance spectrum using the coherence and power spectra following equation 13 of Uttley et al. (2014). This provides the spectral shape of the components which are correlated with the reference band (see e.g. Wilkinson \& Uttley 2009). We use the same segment averaging and reference band as with the time lags, and obtain the covariance spectrum over the 0.5-3.5 mHz frequency range. The coherence between the reference and comparison bands is $>$ 0.4. As expected, the quality of the covariance spectrum varies from observation to observation and is highest in the brightest observations 2, 3 and 7 (see Figure 1).

\begin{figure}
\includegraphics[trim=100 140 50 100, clip, width=8cm]{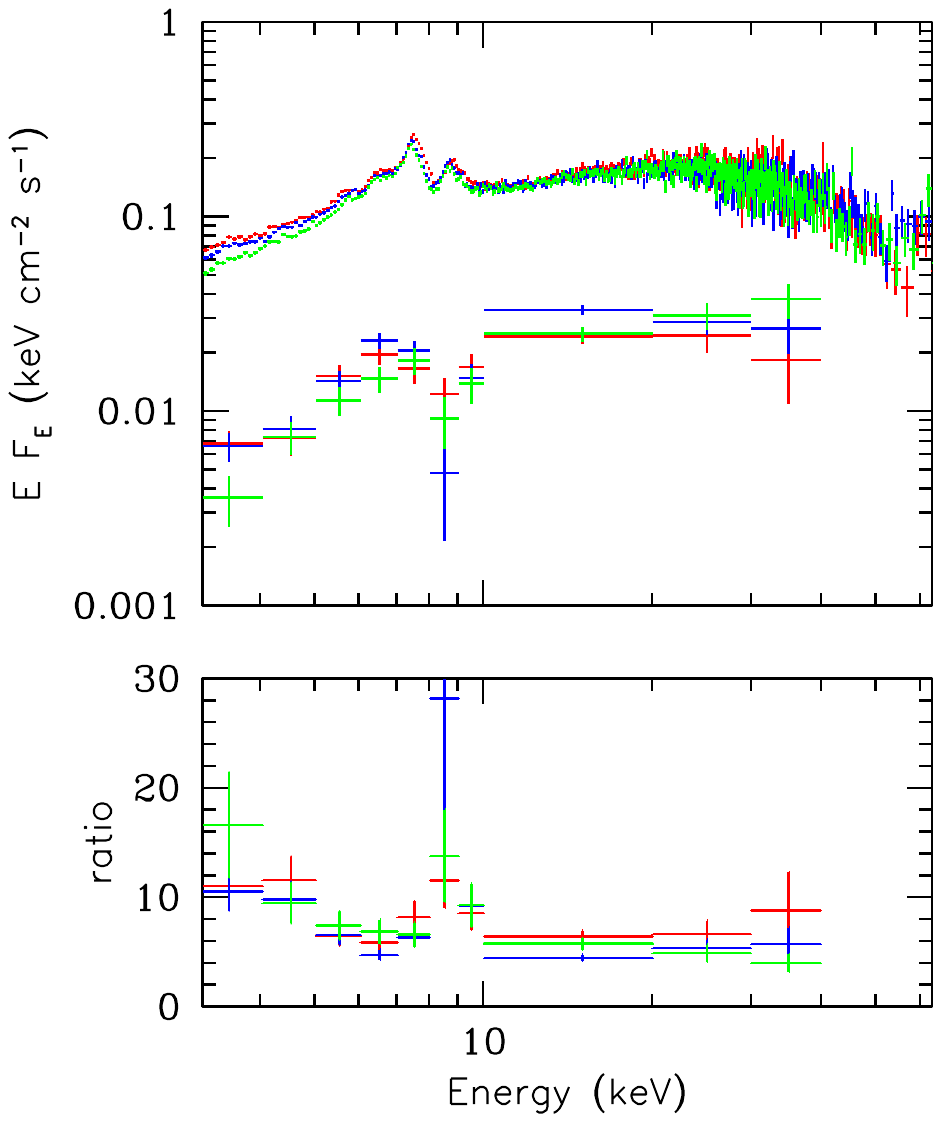}
    \caption{The covariance spectra of Obs 2, 3 and 7 (integrated from 0.5-3.5mHz using the 3-6 keV band as reference) shown in blue, red and green respectively with the time-averaged data (same colours) for the FPMA only, both unfolded through a flat model (power-law of zero index and unity normalization). The ratio of the time-averaged data to the covariance is shown in the lower panel and is inconsistent with a constant, implying that the time-averaged spectrum is composed of multiple components. We note that if the dip in the covariance was associated with the strong emission lines from the jet then we should see a corresponding (and stronger) dip at $\sim$7.5 keV yet this is conspicuously absent.}
    \label{fig:flowchart}
\end{figure}

In Figure 5 we plot the covariance of Obs 2, 3 and 7 along with the time-averaged (FPMA) data from the same observations and the ratio of the two. A constant fit to the ratio indicates a poor description of the data (a null hypothesis probability $<$ 0.05 even when the 8-9 keV bin is excluded) and instead implies that the linearly-correlated variable component in the spectrum does not account for the entire broad-band emission. Attempting to further confirm the shape of this component, we also extract the covariance spectrum when the reference band is at higher energies (formed of the two highest energy bins, i.e. 20-40 keV, to improve the S/N). This is plotted in Figure 6. Due to the poorer data quality and lack of covariance values in the 8-9 keV bin, a constant fit is now acceptable (P $\approx$ 0.12). We note that the ratio of the two sets of covariance spectra (soft to hard reference band) is consistent with unity. It therefore appears unlikely that the shape is being influenced by our choice of reference band.

\begin{figure}
\includegraphics[trim=100 140 50 100, clip, width=8cm]{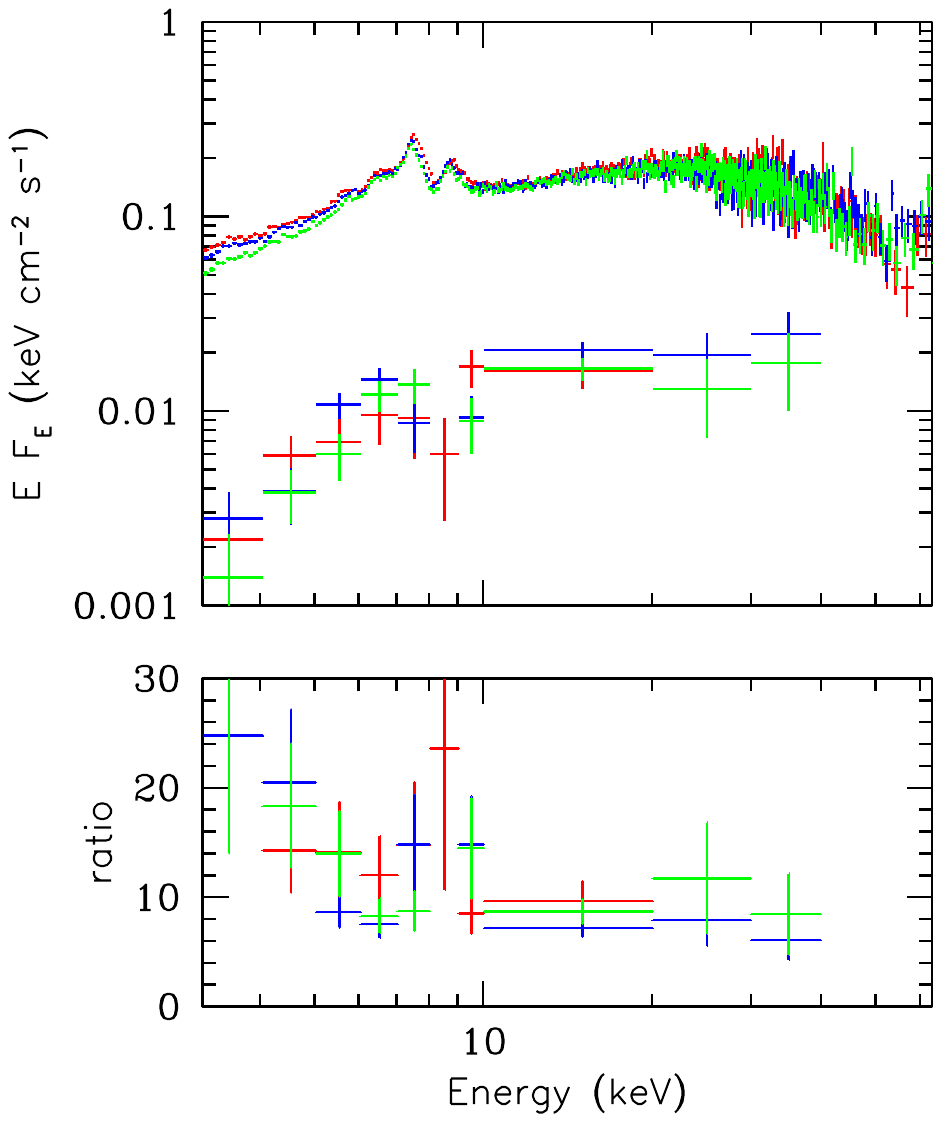}
    \caption{The covariance spectra (obtained using the hard energy reference band of 20-40 keV) of Obs 2, 3 and 7 (integrated from 0.5-3.5mHz) shown in blue, red and green respectively with the time-averaged data (same colours) for the FPMA only, both unfolded through a flat model (power-law of zero index and unity normalization). Although the data is of poorer quality, the shape is clearly similar to that in Figure 5.}
    \label{fig:flowchart}
\end{figure}

Any model for the {\it time-averaged} spectrum of SS433 must also account for the presence and shape of the linearly-correlated variable component. Should the high-energy emission be associated with a hotter component of the jet, then any variability would have to become incoherent over a relatively small distance in order that the covariance is diminished by the time we observe it below 10 keV. Alternatively, we could be observing extrinsic variability imprinted by the crossing of optically thick clumps of wind (Middleton et al. 2011; 2015) covering only the hot part of the jet. However, in order to imprint variability at energies $>$ 20 keV requires Compton thick columns ($>$10$^{24}$ cm$^{-2}$) of material at which point scattering would yield a broad Compton hump and the emergent spectrum is unlikely to be highly variable (as scattering dilutes variability unless the Compton thick wind itself is highly variable). 

Cherepashchuk et al. (2013) explicitly tested a jet origin for the emission at high energies by considering the ratio of the fluxes in the 25-50 keV range seen by {\it INTEGRAL}. By using the Doppler shift inferred from the Fe lines, they 
determined that the changes with precessional phase were inconsistent with relativistic boosting of the approaching jet. This is also problematic if the hard emission is created via Compton scattering of seed photons by electrons in the jet - the emission would be beamed and boosted by the same amount (assuming the ions and electrons are coupled such that the bulk Lorentz factors are the same, and the population of electrons we see are the same at each precessional phase). Given the high throughput and wide bandpass of {\it NuSTAR}, we are able to test this further and ask whether the high energy emission could originate from a portion of the jet which is hotter and faster than the jet which produces the softer emission (and which -- based on our covariance spectra -- would need to vary incoherently). We extract and plot in Figure 7 the count rates in the soft (3-15 keV) and hard (15-60 keV) bands and the hardness ratio (HR = hard/soft) as a function of precessional phase. Whereas precession of a constant velocity jet (emitting both at soft and hard energies) should result in the same amount of Doppler boosting in both bands (with blue shifting of the spectrum enhancing the hardness ratio for lower inclination phases) a faster jet may be able to enhance the change sufficiently to match the factor $\approx$ 2 range in HR we observe. We determine the minimum launch velocity the jet would require in order to match observation, using the standard Doppler boosting formula:

\begin{figure}
\includegraphics[trim=50 170 20 60, clip, width=8cm]{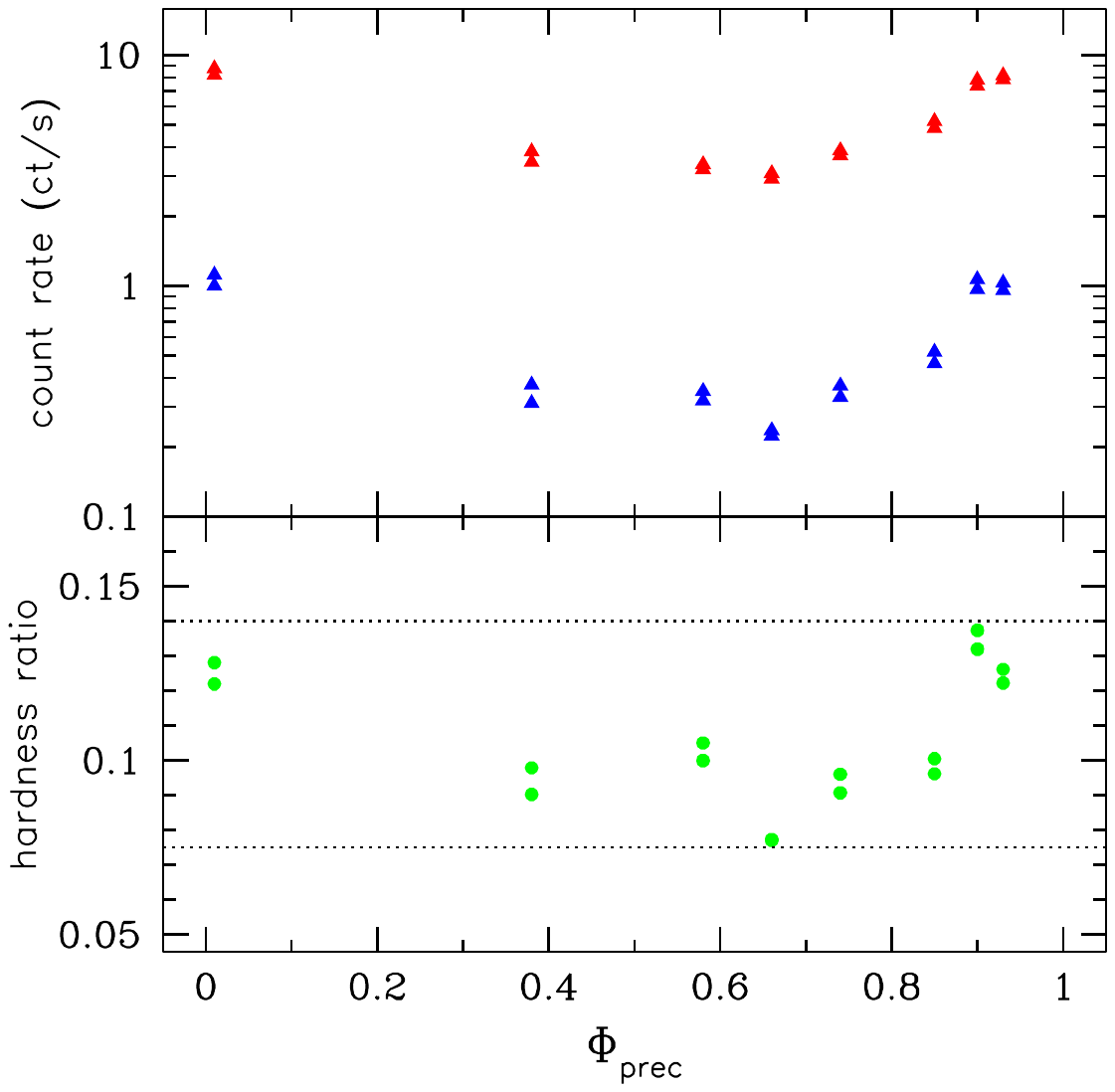}
    \caption{Top panel: soft (red: 3-15 keV) and hard (blue: 15-60 keV) lightcurves for FPMA and FPMB as a function of precessional phase (as determined from the ephemeris - Eikenberry et al. 2001). While the count rate in the soft band extends over a large dynamic range, that in the hard band extends across an even greater range (although the actual count rate itself is far smaller) as can be seen in the bottom panel where the ratio of hard/soft count rates is plotted and which subtends a range of 0.075 to 0.14 (i.e. a factor $\approx$ 2) in hardness ratio over the precession period. In all cases, the statistical errorbars are smaller than the data-points. }
    \label{fig:flowchart}
\end{figure}

\begin{equation}
D = \delta^{3-\alpha} = \left(\Gamma \left[1-\beta cos\left(\theta\right)\right]\right)^{-1\left(3-\alpha\right)}    
\end{equation}

\noindent where $\alpha$ is the index of the spectral component across the energy range of interest, $\Gamma$ is the jet bulk Lorentz factor ($1/\sqrt{\left(1-(v/c)^{2}\right)})$, $\beta = v/c$ (where $v$ is the jet velocity) and $\theta$ is the jet angle to the line-of-sight. Across the precession angles covered by our observations ($\approx$ 58-90 degrees -- Table 1) we see a change in boosting (HR$_{\theta 1}$/ HR$_{\theta 2}$) of at most a factor 2 (Figure 7). Thus, we have a relationship between HR, the cold (soft X-ray emitting) jet (subscript c) and hot (hard X-ray emitting) jet (subscript h) as a function of inclination (between $\theta_{1}$ and $\theta_{2}$):

\begin{align}
\frac{HR_{\theta 1}}{HR_{\theta 2}} = \frac{D_{h1}S_{h}/D_{c1}S_{c}}  {D_{h2}S_{h}/D_{c2}S_{c}} = \left[\frac{\left(\delta_{h}/\delta_{c}\right)_{\theta_{1}}}{\left(\delta_{h}/\delta_{c}\right)_{\theta_{2}}}\right]^{3-\alpha} \notag \\
 = \left(\frac{\left[1 - \beta_{c}cos(\theta_{1})\right]/\Gamma_{h1}\left[1-\beta_{h}cos(\theta_{1})\right]}{\left[1 - \beta_{c}cos(\theta_{2})\right]/\Gamma_{h2}\left[1-\beta_{h}cos(\theta_{2})\right]}\right)^{3-\alpha}
\end{align}

\noindent where $S$ denotes intrinsic (i.e. un-boosted) flux for each component. While the bulk Lorentz factors for the cold jet ($\Gamma_{c}$) have cancelled in the above formula (as the jet component emitting the Fe lines has a well-established, even if sometimes variable, velocity - Blundell \& Bowler 2005), those for the hot jet ($\Gamma_{h}$) do not, as the model we wish to test would predict that we see a {\it different} hot jet velocity at different inclinations to the system. We make a crude approximation for this function:   

\begin{equation}
\beta_{h} = \frac{\left(v_{max} - v_{c}\right)cos(\theta)}{c} + \beta_{c}    
\end{equation}

\noindent where $v_{max}$ is the jet launch velocity and $v_{c} \approx$ 0.26c (i.e. $\beta c \approx$ 0.26). We can numerically solve for $v_{max}$ knowing that at most $\theta_{1}$ = 58 degrees, $\theta_{2}$ = 90 degrees and assuming a reasonable lower limit on the spectral index of the thermal Bremsstrahlung emission in any broad energy band of $\alpha$ = -0.7 (which is the equivalent to $\Gamma$ = 1.7 for a Comptonised power-law component). We find that, in order to match the observed dynamic range in hardness ratio, there is no acceptable solution for $v_{max}$, i.e. unphysically large velocities of $>$0.999c are required. 

The most viable explanation for the covariance spectra and relative changes in brightness as a function of precessional phase therefore appears to be a changing view of emission from within the wind cone -- be it reflected intrinsic flux by the precessing optically thick wind-cone (e.g. Cherepashchuk et al. 2005) or Comptonisation by plasma interior to the wind cone (see Cherepashchuk et al. 2020 and discussions therein). We note that the amount of fractional variability seen in the linearly correlated component at high energies is 20-30\% which, by comparison with other accreting binaries (including both X-ray binaries: Munoz-Darias et al. 2011 and ULXs: Sutton et al. 2013; Middleton et al. 2015), implies that the entire hard emission in SS433 is dominated by a single variable component, the implications of which we proceed to explore in the following sections.

\begin{table*}
\begin{center}
\begin{minipage}{170mm}
\bigskip
\caption{Best-fitting parameters for the fits to the covariance spectra.}
\begin{tabular}{l|c|c|c|c}
\hline 
\hline
Model component 
(parameter | units)   &  {\sc tbabs*pexmon}    &  {\sc tbabs*gabs*pexmon}    &  {\sc tbabs*gabs*bremss} & {\sc tbabs*gabs*comptt}\\
\hline
{\sc tbabs} (n$_{\rm H}$ | cm$^{-2}$) & $<$ 1.49 $\times$10$^{22}$   &  $<$ 3.13 $\times$10$^{22}$ & 18.14 $^{+3.36}_{-3.11}$ $\times$10$^{22}$ & $<$ 11.00 $\times$10$^{22}$\\

{\sc pexmon} (E$_{\rm fold}$ | keV )& $>$ 46.30  & $>$ 37.81  &  ----- & ----- \\

{\sc pexmon} ($\Gamma$)&  2.24 $\pm 0.06$  &  2.36 $^{+0.11}_{-0.08}$ &  ----- & -----\\

{\sc pexmon} (z)&  -7.62 $^{+1.12}_{-0.84}\times$10$^{-2}$  &  -9.89 $^{+0.86}_{-0.79}\times$10$^{-2}$ &  ----- & -----\\ 

{\sc pexmon} (Fe | solar abundance) & 0.85 $\pm 0.13$   &  0.66 $^{+0.18}_{-0.15}$  &  ----- & -----\\

{\sc bremss} (kT | keV) & ----- & ----- & 28.30 $^{+4.94}_{-3.92}$ & ----- \\

{\sc comptt} (kT$_{\rm e}$ | keV) & ----- & ----- & ----- & $>$ 9.37\\

{\sc comptt} ($\tau$) & ----- & ----- & ----- & $<$ 3.36\\

{\sc gabs}  ($\sigma$ | keV) &  -----  & 0.69 $^{+0.22}_{-0.15}$  & 0.45 $^{+0.13}_{-0.14}$ & 0.55 $^{+0.11}_{-0.10}$\\

{\sc gabs}  (strength)  & -----   & 1.53 $\pm 0.33$ & 1.25 $^{+0.31}_{-0.27}$ & 1.62 $^{+0.32}_{-0.29}$\\

$\chi^{2}$ (d.o.f.)  &  60 (21)  & 29 (19)  & 50 (23) & 45 (21)\\
\hline
\hline
\end{tabular}

Notes: Spectral models and best-fitting parameters (with 1$\sigma$ errors) from fitting to the covariance spectra shown in Figure 5. 
\end{minipage} 
\end{center}
\end{table*}

\subsubsection{Covariance spectral fitting}

Based on the previous arguments, we fit the highest-quality covariance spectra (Obs 2, 3 \& 7) with a model for absorbed reflection and a model for Comptonisation. We choose a simple model to fit the data in {\sc xspec v12.10.1} of {\sc tbabs*pexmon} where {\sc pexmon} describes reflection of an incident cut-off power-law from neutral material and has self-consistent Fe and Ni lines (Nandra et al. 2007), and {\sc tbabs} accounts for absorption by neutral intervening material (with abundances from Wilms et al. 2000) for which we set the lower limit to be the line-of-sight column in the direction of SS433 (7$\times$ 10$^{21}$ cm$^{-2}$: Dickey \& Lockman 1990; Kalberla et al. 2005). We set the reflection scaling factor to be negative such that only the reflected emission is observed (i.e. the intrinsic continuum is absent) and add a multiplicative constant offset between observations. The resulting best-fit model parameters are presented in Table 2. We proceed to include a multiplicative {\sc gabs} component with a centroid energy fixed at 8.5 keV (to approximate the center of the bin in which the lag is detected). The inclusion of this component results in a significant change in $\Delta\chi^{2}$ (31 for 2 extra d.o.f.). Although $\Delta\chi^{2}$ can be misleading when including additional components (Protassov et al. 2002), in this case, the lag spectra provide an additional, compelling reason to include such a line in our subsequent spectral fitting.

We note that the Doppler shift of the reflection component in our best-fitting model implies motion towards the observer and at the 3$\sigma$ level is $<$ 0. Although the emission we observe will no-doubt be a complex weighting of reflection from material moving with a range of velocities to the line-of-sight, we note that the inferred motion is degenerate with the ionisation state of the reflecting material. Indeed, it is highly probable -- given the high accretion rates in this source and the radiation field which is obscured from view -- that the material will be at least partially ionised rather than neutral (see Medvedev \& Fabrika 2010, and Pinto et al. 2020 for a discussion on the thermal stability of such winds) as we have assumed for the sake of simplicity. In the future we will utilise bespoke reflection models built to explore this geometry. We also note that the best-fitting Fe abundance appears to be sub-solar; this is interesting in light of the clearly super-solar abundance in the jet (Marshall et al. 2013) although may again be a consequence of our simplified model. Finally, although we allowed the inclination of the reflector to be a free parameter, the value is unconstrained at 3$\sigma$.

To test whether the hard component could instead be described by thermal Comptonisation, we fit the covariance spectra using a model of {\sc tbabs*gabs*comptt} (see Titarchuk 1994 for a description of {\sc comptt}). The fit implies a relatively hot, relatively low optical depth ($\tau \lesssim 3$) plasma; however the fit quality is significantly worse than with the reflection model (a difference of $\Delta\chi^{2}$ of 16 for a change of 2 d.o.f.). As such, we do not utilise Comptonisation in the following modelling but refer to the implications of a corona in the Discussion \& Conclusions.

Although we have a strong physical argument against the jet forming the hard energy emission in SS433 -- be it a hotter component or as a site for Comptonisation (assuming the electrons do not cool significantly in the time taken to traverse the depth of the wind-cone) -- for completeness we also attempt to fit the covariance spectrum using a simple model for the jet, i.e. a Bremsstrahlung continuum. We again include a Gaussian absorption line based on the time-lag, such that the total model is now {\sc tbabs*gabs*bremss}. However, as indicated in Table 2, we find that in order to fit the data, the neutral column density has to be over an order of magnitude larger than reported in other studies ($\sim$10$^{23}$ versus $\sim$10$^{22}$cm$^{-2}$, Medvedev \& Fabrika 2010). If we fix the column density to either $\approx$ 5$\times$10$^{22}$ cm$^{-2}$ (found here in our fits to the time-averaged data - see section 2.4) or 1.5$\times$10$^{22}$ cm$^{-2}$ (reported from spectral fits in the literature, Medvedev \& Fabrika 2010), the Bremsstrahlung model gives a much poorer fit to the data than the reflection continuum, with a $\Delta\chi^{2} >$ 40 for 5 additional d.o.f.

\begin{table*}
\begin{center}
\begin{minipage}{120mm}
\bigskip
\caption{Best-fitting parameters for the fits to the time-averaged spectra.}
\begin{tabular}{l|c}
\hline 
\hline
Model component 
(parameter | units)   &  {\sc tbabs*gabs*gsmooth*(gauss + cevmkl +} \\
& {\sc cevmkl + pexmon)}    \\
\hline
{\sc tbabs} (n$_{\rm H}$ | cm$^{-2}$) & 5.77 $^{+0.04}_{-0.05}$ $\times$10$^{22}$     \\
{\sc gabs}  (strength) & $<$ 0.003  \\
{\sc cevmkl} ($\alpha$) & 0.40 $^{+0.04}_{-0.01}$ \\ 
{\sc cevmkl} (Fe | solar abundance) & 2.11 $\pm$ 0.01 \\
{\sc cevmkl} (Ni | solar abundance) & 21.47 $^{+0.19}_{-0.24}$ \\ 
{\sc cevmkl$_{1}$} (kT$_{\rm  max}$ | keV) & 20.42 $^{+0.10}_{-0.09}$ \\ 
{\sc cevmkl$_{2}$} (kT$_{\rm  max}$ | keV) & 21.04 $^{+0.09}_{-0.07}$ \\ 
{\sc cevmkl$_{3}$} (kT$_{\rm  max}$ | keV) & 16.31 $^{+0.04}_{-0.09}$ \\ 
{\sc cevmkl$_{4}$} (kT$_{\rm  max}$ | keV) & 27.17 $^{+0.21}_{-0.16}$ \\ 
{\sc cevmkl$_{5}$} (kT$_{\rm  max}$ | keV) & 17.82 $^{+0.07}_{-0.15}$ \\ 
{\sc cevmkl$_{6}$} (kT$_{\rm  max}$ | keV) & 20.64 $^{+0.08}_{-0.12}$ \\ 
{\sc cevmkl$_{7}$} (kT$_{\rm  max}$ | keV) & 24.24 $^{+0.13}_{-0.12}$ \\ 
{\sc cevmkl$_{8}$} (kT$_{\rm  max}$ | keV) & 26.79 $^{+0.14}_{-0.16}$ \\ 
{\sc pexmon} ($\Gamma$) & 1.38 $\pm$ 0.01 \\ 
{\sc pexmon} (E$_{\rm fold}$ | keV )  & 27.39 $^{+0.27}_{-0.17}$ \\ 
$\chi^{2}$ (d.o.f.) & 6081~~(4529)       \\
\hline
\hline
\end{tabular}

Notes: Spectral model and best-fitting parameters (with 1$\sigma$ errors) from fits to the time-averaged spectra shown in Figure 8. 
\end{minipage} 
\end{center}
\end{table*}

\subsection{Time-averaged spectral fitting}

\begin{figure*}
\includegraphics[trim=20 180 20 60, clip,width=12cm]{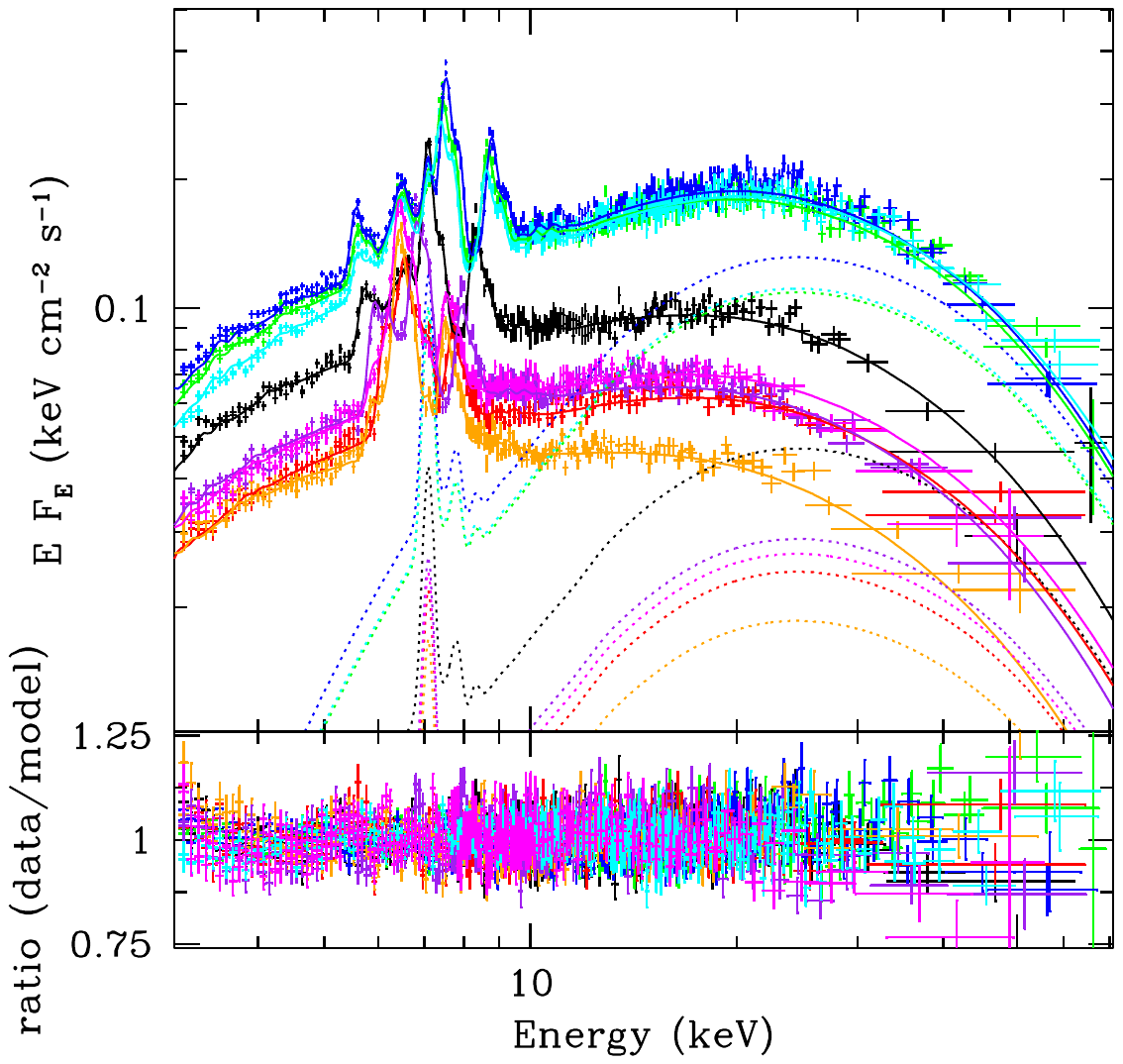}
    \caption{Time-averaged data for all observations (with data from both {\it NuSTAR} detectors) unfolded through the best-fitting spectral model. The reflection component ({\sc pexmon}) is highlighted as a dotted-line from which we determine the apparent luminosity ($L_{\rm ref}$) as a function of inclination as presented in Table 1. The lower panel shows the ratio of data to model.}
    \label{fig:flowchart}
\end{figure*}

Under the reasonable assumption that reflection may dominate the hard X-ray emission of SS433, we create a model for the time-averaged spectrum. The simplest possible model we can adopt for the jet is free-free emission with collisionally excited emission lines and a weighted emission measure: $dEM = (T/T_{\rm max})^{\alpha-1} dT/dT_{\rm max}$ ({\sc cevmkl}: Singh et al. 1996 with {\sc switch} = 2 such that the model utilises {\sc atomDB}). The weighting accounts for the multi-temperature nature of the jet and we allow each observation to have a free $T_{\rm max}$ but a tied index ($\alpha$). We include two {\sc cevmkl} components for the approaching and receding jets respectively, with free abundances of Fe and Ni (the most relevant for our bandpass and, in the case of Ni known to be highly super-solar: Marshall et al. 2013). We tie the plasma temperature in the approaching and receding jets for each observation, consistent with previous studies (Marshall et al. 2002). We also include reflection ({\sc pexmon}, as described above) with electron temperature and photon index tied across all observations. In doing so we are making the explicit assumption that the irradiating continuum does not change greatly with time. We initially fix the Fe abundance in the reflecting material to that indicated by fits to the covariance spectrum (and explore the effect of this later).

We convolve the above model components with a Gaussian smoothing profile ({\sc gsmooth} with a width equal to $\sigma (E/6$ keV$)^{\alpha}$ and $\alpha$ = 1) to account for broadening due to the opening angle of the jet and any additional Comptonisation/turbulent motion (Fabrika 2004), and presume that this is not dissimilar to any velocity broadening by the optically thick material in the wind cone. We also include a Gaussian absorption line ({\sc gabs}) at 8.5 keV with the best-fitting line-width fixed to that from the fits to the covariance spectra (but normalisation left free to vary), and absorption by neutral gas in the line-of-sight to SS433 ({\sc tbabs}). We also include a stationary Fe K$_{\alpha}$ line at 6.4 keV (as reported in Kotani et al. 1996 and Kubota et al. 2010) and tie the inclinations of the reflector between observations at similar precessional phases ([1, 2, 3 \& 7] and [4, 5, 6 \& 8]). Finally, we apply a constant offset between detectors to account for any differences in instrument response (typically $<$ 5\%, Madsen et al. 2015). The model in {\sc xspec} is: {\sc tbabs*gabs*gsmooth*(gauss + cevmkl + cevmkl + pexmon)}. We proceed to fit all spectral datasets (both FPMA and FPMB) simultaneously with the above model and {\sc cevmkl} plasma temperatures free to vary between observations. 

The best-fitting model yields a reasonably good reduced $\chi^{2}$ = 1.34 (Table 3), however, the corresponding null hypothesis probability is $\ll$ 0.05 (and therefore statistically rejectable) due to the extremely high data quality, the probable impact of additional line structure (Marshall et al. 2002), and as a consequence of tying continuum parameters across observations (whereas in reality these will vary over time to some degree). However, as can be seen in Figure 8, in all cases, the strongest line features and the continuum are well-described, which is crucial for our analysis. The best-fitting model parameters of interest are provided in Table 3 with their 1$\sigma$ errors. We note that the abundance of both Fe and Ni are found to be super-solar as reported in previous campaigns (Marshall et al. 2013), with a similar ratio of Ni/Fe ($\approx$ 10) and absolute values for the abundance, to those reported in other studies (see Khabibullin, Medvedev \& Sazonov 2016). The temperatures of the {\sc cevmkl} component range from $\approx$ 16-27~keV without any obvious correlation with precessional phase. In Figure 9 we plot the best-fitting Doppler shifts of the approaching and receding jets versus precessional phase alongside the model predictions (Eikenberry et al. 2001). In most cases, the values are consistent with expectation except at the cross-over phases ($\Phi_{\rm prec}$ = 0.38 and 0.66, where it is likely that the fit only requires a single jet component). To test the impact on our results, we freeze the receding and approaching Doppler shifts to their predicted values at those phases, finding the flux in the reflected component to change by at most 9\% but typically by $<$ 1\%.  

\begin{figure}
\includegraphics[trim=200 100 200 100, clip, width=9cm]{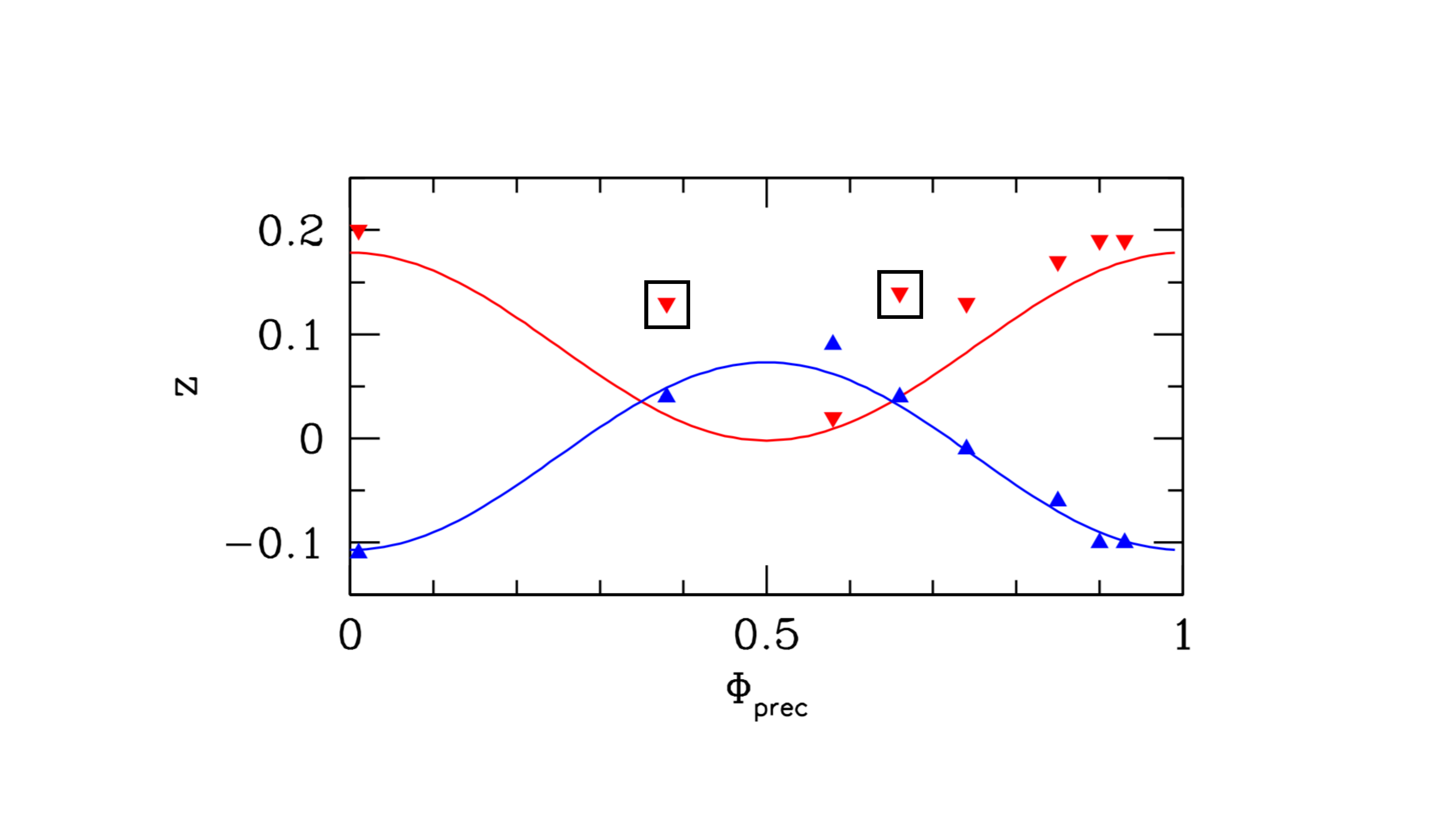}
    \caption{Doppler shifts of the approaching (blue) and receding (red) jets from our best-fitting model. At the cross-over phases the points deviate from the model as indicated by the boxed in points (though fixing the Doppler shifts in those cases does not influence our findings).}
    \label{fig:flowchart}
\end{figure}

We also note that the $\sigma$ component of the {\sc gsmooth} model, tied across all observations, is $<$ 0.1~keV, and the inclination of the reflecting plasma (tied across observations [1, 2, 3 \& 7] and [4, 5, 6 \& 8], was found to be $> 84$ degrees (at the 1$\sigma$ level, noting that the model is limited to inclinations $<$ 85 degrees). This latter result is somewhat surprising given we expect to observe at lower inclinations to the far side of the wind-cone. To test the impact of this, we fix the inclination for all observations to zero degrees (i.e. face-on to the reflector), finding the overall fit quality to be slightly worse ($\chi^{2}$ = 6176 for 4531 d.o.f.), with all of the fluxes in the reflection component dropping by a factor $< 2.1$ compared to the previous best fit. As we will see, the final inferred lower limits on the apparent luminosity are sufficiently high that this will not adversely affect the overall findings but this will be fully explored when a more geometrically appropriate reflection model becomes available.  

It is interesting to note that the spectral index ($\Gamma$) of the {\sc pexmon} component is considerably harder than we found in fits to the covariance spectrum, but is consistent with values inferred for the continuum of some ULXs, albeit with a higher cut-off energy than has been observed (see e.g. Pintore et al. 2017, Walton et al. 2018). There may well be a physical reason for the covariance requiring a softer irradiating spectrum if the variable component of the reflection originates from a softer irradiating power-law (e.g. emission from outside of the innermost regions). This will be explored in the creation of new spectral models which more accurately describe these geometries. For now we explore the impact by setting the spectral index in the time-averaged fits to be fixed at $\Gamma$ = 2.36 (see Table 2); the resulting fit is considerably worse ($\chi^{2}$ = 8734 for 4530 d.o.f.) but the fluxes from the reflection component remain within 15\% of the previous values. 

We also explore the impact of tying the Fe abundance between the jet and reflection components, finding the fit to be considerably poorer ($\chi^{2}$ = 7711 for 4529 d.o.f.) with an Fe abundance of 2.06 $\pm$ 0.01 and all 
reflected fluxes reduced by factors of $<$ 2.1. Finally, we allow the Fe abundance in the reflection component to be free, finding a preferred fit ($\chi^{2}$ = 6012 for 4528 d.o.f.) with a lower value of Z = 0.44 and slightly higher reflected fluxes (but within 25\% of those in our best-fit model with a frozen Fe abundance). We stress that these numbers should be treated with caution, as the model is as yet only a crude approximation to the true reflecting plasma (both in terms of geometry, density and ionisation state).

\begin{table*}
\begin{center}
\begin{minipage}{107mm}
\bigskip
\caption{Intrinsic and apparent luminosities of SS433.}
\begin{tabular}{l|c|c|c|c}
\hline 
\hline
  &    \multicolumn{4}{c} {inclination (observation)}\\    
  &  68 (1) &  60 (2) & 59 (3) & 63 (7) \\
  \hline
$L_{\rm obs,35}$ (erg/s) & 5.20 & 12.05 & 14.56 & 12.30 \\
Correction factor [0.15/0.3c] & 0.018/0.009 & 0.042/0.025 & 0.046/0.026 & 0.030/0.018 \\
$L_{\rm int,37}$ (erg/s) [0.15/0.3c] & 2.91/5.47 & 2.86/4.77 & 3.15/5.50 & 4.08/6.95 \\
$L_{\rm app,39}$ (erg/s) [0.15/0.3c] & 1.28/2.52 & 1.26/2.19 & 1.38/2.53 & 1.80/3.20 \\
\hline
\hline
\end{tabular}

Notes: luminosities of the observed reflected component in SS433 (in units of 10$^{35}$ erg/s), the inferred intrinsic, irradiating luminosity (in units of 10$^{37}$ erg/s), and observed face-on luminosity (in units of 10$^{39}$ erg/s), based on the correction factors provided in the table for outflow velocities of 0.15c and 0.3c, wind-cone half opening angle of 10 degrees and height of 100,000 R$_{\rm g}$. These are robust lower limits given the high likelihood of a smaller wind-cone opening angle, advection along the jet, and more complicated reflection modeling, all of which are expected to increase the intrinsic and apparent luminosities.
\end{minipage} 
\end{center}
\end{table*}

\section{SS433 as a ULX}

\begin{figure}
\includegraphics[trim=100 260 10 100, clip, width=9cm]{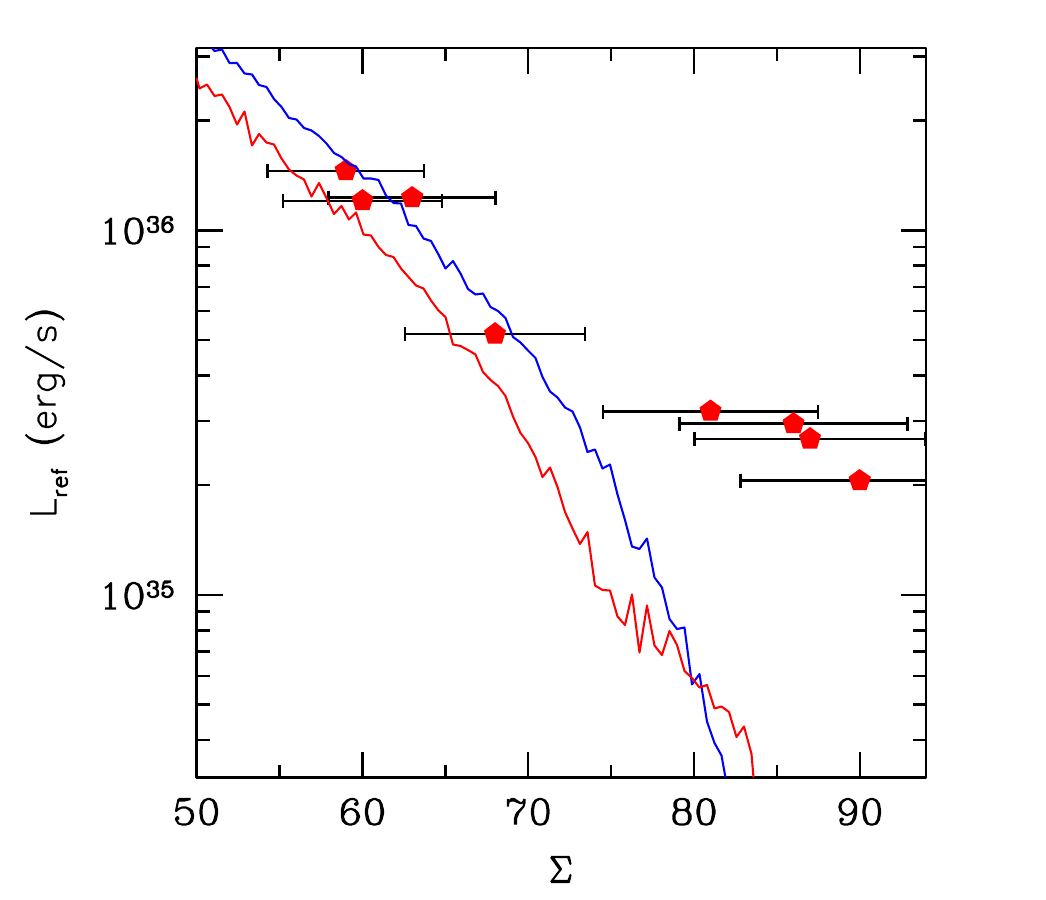}
    \caption{De-absorbed (3-60 keV) reflected luminosities for each observation as a function of inclination. The errors on the luminosity are smaller than the data points and the error on the inclination is set at 8\% of the inferred inclination (a conservative upper limit on the true uncertainty - Eikenberry et al. 2001). The plotted lines are theoretical curves for a 10 degree half-opening angle of a wind-cone (Dauser et al. 2017), with material outflowing at 0.15c (blue) and 0.3c (red). While the curves are not fitted to the data, they have been offset vertically for the sake of clarity. Above $\approx$ 70 degrees the data clearly departs from the model predictions, likely due to flux escaping through less dense regions of the wind.}
    \label{fig:flowchart}
\end{figure}

From our best-fitting spectral model (with the Fe abundance fixed to that from the covariance spectral fits), we extract the integrated, de-absorbed reflected flux across 3-60 keV by applying a {\sc cflux} convolution model to the {\sc pexmon} component. From the fluxes, we derive the unabsorbed luminosities shown in Table 1 from the assumed distance to the source of $\approx$ 6 kpc (Lockman et al. 2007) and errors from the propagated uncertainty on the flux. 

We note again that our choice of a neutral reflector in our spectral modelling ({\sc pexmon}) is a clear over-simplification and, in reality, the reflecting material is likely to have multiple ionisation zones. As we are simply determining the flux from integration of the model, our choice of a simple, neutral reflector rather than a more complex, partially ionised model, is unlikely to have any major impact on the result (as the reflected flux from a neutral model is naturally smaller than from a partially ionised one). This simplification will be directly addressed through the development and application of new models.

Using the well-established kinematic model (see Eikenberry et al. 2001) we determine the inclination of the jets -- and we assume therefore the wind-cone -- to our line-of-sight ($\Sigma$):

\begin{equation*}
    {\rm cos}(\Sigma) = {\rm cos}(2\pi \Phi_{\rm prec}){\rm sin}(\theta){\rm sin}(i) + {\rm cos}(\theta){\rm cos}(i)
\end{equation*}

\noindent Where $i$ is the mean system inclination of 78.8 degrees (Margon \& Anderson 1989) and $\theta$ is the precession cone half-angle, known to be $\approx$ 20 degrees (Eikenberry et al. 2001). From the ephemeris of Eikenberry et al. (2001), we determine the inclination of the approaching jet to our line-of-sight, list these in Table 1, and plot them against the corresponding reflected luminosity in Figure 10. Although highly unlikely, assuming that jitter in the Doppler shift of the optical jets (Eikenberry et al. 2001) results entirely from jitter in the inclination, we can place an upper limit on the error for the inclination of $\approx$ 8\%. 

A numerical approximation to geometrical beaming has been obtained by performing a series of Monte-Carlo simulations of photons achromatically scattering a number of times until escaping from an optically thick wind-cone (Dauser et al. 2017, see also Abolmasov 2009). This approach allows us to take an observed reflected luminosity for a given inclination and obtain an estimate for the intrinsic source luminosity (assumed to be located at the base of the wind-cone), for a wind with a given opening angle and outflow velocity. In Figure 10 we overlay two of these theoretical models for half-opening angles of 10 degrees (a conservative underestimate for the opening angle given the enormous accretion rates - Jiang et al. 2019), a wind-cone height of 100,000 $R_{\rm g}$ ($R_{\rm g} = GM/c^{2}$, where $G$ is the gravitational constant, $M$ is the mass of the compact object and $c$ is the speed of light), and outflow velocities of 0.15c and 0.3c (as inferred from the lag spectrum) respectively onto our data. The wind-cone height was set by computational restrictions and represents an under-estimate of the true wind-cone height in this source (Marshall et al. 2002). We stress that we have simply scaled these models, they are not fit to the data. We also note that for those inclinations $>$ 90 degrees, we have translated their value to 180 - $\Sigma$ as the beaming model is symmetrical.

A good description of all the data is clearly not yet possible which is unsurprising in light of the simplistic nature of the model (in future it may be possible to extract equivalent values directly from 3DRMHD models, e.g. Narayan et al. 2017). The data appears to be reasonably described up to $\approx$ 70 degrees and then over-shoots the model at inclinations above this. This is to be expected due to the limitations inherent in treating the wind cone as impermeable -- indeed we know that some X-ray flux must escape through the wind in order for absorption features to be imprinted. From simple geometrical considerations, we expect this discrepancy to be most obvious at high inclinations. Assuming the model is a reasonable description for inclinations below 70 degrees, we obtain the correction factors to the observed fluxes given in Table 4. From these we then obtain a measure of the intrinsic flux (from flux$_{\rm obs}$/correction factor) which we find to be $\gtrsim$ 2$\times$10$^{37}$ erg/s. This is likely to be a lower limit, as some flux may be advected along the jet or lost through Compton scattering in the wind (see Fabrika 2004). Incorporating the beaming factor for inclinations less than the 10 degrees half-opening angle of the wind-cone (with factors lying between 44-50 and 46-52 for $v_{\rm wind}$ of 0.15 and 0.3c respectively) we then estimate the apparent X-ray luminosity a face-on observer would infer, to be $>$ 1$\times$10$^{39}$ erg/s. We expect this to be a stringent lower limit should the wind cone be more narrow (as expected at these accretion rates - see Jiang et al. 2019) or be of greater height. We can be confident that regardless of the uncertainties, the apparent luminosity determined by a face-on observer would qualify SS433 as a ULX.

\section{Discussion \& conclusions}

{\it NuSTAR} observed SS433 in the 3-79 keV range, sampling its 162 day precessional period, allowing for the first, high throughput, time-resolved analysis at high energies where the intrinsic emission from the source is thought to emerge. 

By comparison to the time-averaged data, the covariance indicates that a two-component spectrum is required. The fractional covariance (analogous to the fractional rms) at high energies is 20-30\%, consistent with the largest amounts of variability typically detected in accreting sources (Munoz-Darias et al. 2011), implying that the entire hard excess is formed from this linearly correlated component. The shape of the covariance spectrum shown in Figure 5 best matches that of reflection from optically thick material, with a pronounced drop, consistent with an Fe K photoelectric edge (albeit blue-shifted) and/or absorption lines, and clearly not associated with the jet emission lines. 

To better understand the spectrum of SS433, we also extracted frequency and energy-dependent time-lags. As shown in Figures 2 and 3, we detect the presence of a highly significant lag at 8-9.3 keV in several observations, varying in magnitude with frequency. A lag across such a narrow energy range implies an atomic (rather than continuum) process. 
The absence of multiple features would argue against the lag being connected to atomic emission associated with the jet. Instead, the lag is most probably imprinted by either transmission through outflowing, partially ionised gas with the magnitude of 10s - 100s of seconds, providing an indication of the combined distance and density of the intervening material (Middleton et al. 2019; see also Kara et al. 2020 for a similar sized lag found recently in NGC1313 X-1), recombination of the plasma (Silva et al. 2016) due to changing irradiation, or some combination of the two effects. It is unlikely that the lag is associated with a photoelectric edge in a thin jet sheath due to the implied line-of-sight velocities and small opening angle of the jet -- although it is certainly possible that some imprint of an edge could be present in the spectrum, e.g. Kubota et al. (2007).

Given the energy of the lag feature (which can extend up to 9.3 keV - Figure 3), we consider that the line could be Fe XXV or Fe XXVI, blue-shifted due to the line-of-sight velocity of the outflow, which may be depleted in Nickel relative to the jet (Medvedev et al. 2018). We consider this the more likely explanation compared to absorption by Nickel, as, in the latter case, lines would be expected from both ionised species of Nickel and Iron (Figure 4), yet only one is present. Assuming the line is either Fe XXV or Fe XXVI, we find a lower limit on the outflow velocity of 0.14-0.29c/0.28-0.32c (which is a lower limit as we may not be viewing along the direction of the outflow). Should the feature in the lag spectrum instead be associated with an Fe K edge via propagation (requiring a velocity gradient) then we would infer similar line-of-sight velocities (up to 0.26c). These velocities are consistent with outflows seen in simulations (e.g. Jiang et al. 2014) and detected in the X-ray spectra of ULXs (Pinto et al. 2016, 2017; Walton et al. 2016, Kosec et al. 2018), also known to harbour super-critical accretion flows. Such line-of-sight velocities are in-excess-of those seen in the approaching jet ejecta at any precessional phase (Eikenberry et al. 2001), ruling out an origin in this component. In future we will directly model the lag spectra to better constrain the nature of the wind (considering propagation and recombination timescales).

To determine the reflected flux as a function of precessional phase, we fit the time-averaged data with a model composed of reflection and a model for the jet at softer energies. We convert the reflected fluxes into inclination-dependent luminosities using the well-established kinematic model (Eikenberry et al. 2001) and distance to SS433 of $\approx$~6 kpc (Lockman et al. 2007). The observed luminosities can then be compared to theoretical models for reflection inside an optically thick wind-cone (Dauser et al. 2017), and a lower limit on the intrinsic X-ray luminosity extrapolated.

Our theoretical model assumes a simple conical geometry with achromatic scattering, and allows us to infer a `correction factor' to translate the observed into the intrinsic luminosity. We assumed a wind velocity of 0.15c - 0.3c to be consistent with the absorption feature revealed by the lag, a wind-cone half opening angle of 10 degrees (a conservative underestimate for such high accretion rates - Jiang et al. 2019) and a wind-cone height of 100,000 R$_{\rm g}$. The latter is likely to be a conservative underestimate given the distance from the cone's apex at which we detect the hottest jet emission, and is expected to be up to around an order of magnitude larger (Marshall et al. 2002). These input values allow us to construct the models shown in Figure 10. Whilst the tail of the theoretical curve (at high inclinations) deviates strongly from the data, this may occur due to flux leaking out through the wind. Although simplistic, the description of the data by the model appears to be reasonable up to $\approx$70 degrees.

Assuming the correction factors for inclinations below 70 degrees, we infer a lower limit on the intrinsic X-ray luminosity for SS433 of $\ge$ 2$\times$10$^{37}$ erg/s. This is necessarily a lower limit as some portion of the intrinsic X-ray flux will likely be down-scattered or absorbed by the wind (e.g. Fabrika 2004). While we only observe a faint apparent X-ray source, a {\it face-on} observer (e.g. in a local galaxy) would see an amplified luminosity as a consequence of collimation by the wind-cone (King et al. 2001; 2009). For a half-opening angle of 10 degrees, the X-ray luminosity would appear to be $>$ 1 $\times$10$^{39}$ erg/s, and SS433 would be classified as a ULX. This is a robust lower limit; should the wind-cone have a greater height as implied by observations of the jet plasma (Marshall et al. 2013) and be narrower as implied by simulations at the extremely high accretion rates of SS433 (Jiang et al. 2019), then we would expect even greater amplification of the face-on luminosity.

It has been suggested that the hard excess in SS433 could instead be described as Compton up-scattering by optically thin plasma in the wind cone (e.g. Cherepashchuk et al. 2005, 2007, 2013, 2020; Krivosheyev et al. 2009), perhaps generated as a result of collisions between the wind and jet (e.g. Begelman et al. 2006). Whilst we find that the covariance spectrum is better described by reflection rather than Comptonisation, should up-scattering take place in a complex environment with a range of densities, temperatures and velocities, it is not clear what shape the spectrum of the linearly correlated variability may take. Studying the details of such a corona is beyond the scope of this paper, however, we note that, as the corona is expected to be optically thin, the effect on the geometrical beaming/amplification may be minimal (in an optically thin plasma a photon has a higher probability of scattering off the walls of the wind than being Compton up-scattered). We also note that both reflection and coronal models reproduce the observation that hard X-ray eclipses are longer in duration than at soft energies (Kawai et al. 1989; Yuan et al. 1995; Cherepashchuk et al. 2020 -- a natural result of the hard emitting region subtending a larger area in the path of the secondary star than the jet). In future it may be possible to distinguish between these models through observations covering the eclipse, should the corona be expected to be quasi-static (for instance created in a shock) rather than quasi-Keplerian as the wind should be.

SS433 has long been suspected to be a ULX if seen face-on (e.g. King et al. 2001; Begelman et al. 2006; Fabrika et al. 2015; Khabibullin \& Sazonov 2016, Waisberg et al. 2018) and our constraint on the intrinsic and apparent X-ray luminosity fully supports these assertions. Although the mean and precessional inclinations of SS433 are well established, this is not the case for any other ULXs. In a small number of cases, eclipsing ULXs (e.g. Urquhart \& Soria 2016) indicate high system inclinations, although the wide-spread phenomenon of precession (due to an as-yet unknown mechanism, see Mushtukov et al. 2017; Middleton et al. 2015, 2018; Vasilopoulos et al. 2019) can complicate our understanding of the source orientation, especially if driven by misalignment between the binary and compact object spin axis (Middleton et al. 2019; King \& Lasota 2020). The situation is complicated further as, unlike SS433, should a given ULX have a low enough Eddington-scaled accretion rate such that the wind is of limited optical depth (Poutanen et al. 2007), or if the compact object is a highly magnetized neutron star such that the X-ray emission emerges quasi-isotropically from an accretion curtain (Mushtukov et al. 2017) at radii larger than the spherisation radius (e.g. Middleton et al. 2019), then it could appear X-ray bright whilst still being viewed at high inclination. Regardless of the complications, SS433's ULX nature demands that many such systems will be present throughout the universe yet will not be considered as ULXs (see Wiktorowicz et al. 2019), having an X-ray luminosity below 10$^{39}$ erg/s (e.g. Middleton et al. 2015) or not being visible in the X-rays at all (although these may yet be detected by their bubble nebulae: Soria et al. 2014, 2020).

We note that our constraint on the intrinsic X-ray luminosity of SS433 ($>$ 2$\times$10$^{37}$ erg/s) appears around or more than an order of magnitude smaller than the kinetic luminosity of the jet (10$^{38}$ - 10$^{39}$ erg/s: Marshall et al. 2002). The implication is that either there is an additional launching mechanism (other than radiation pressure) contributing to the mechanical jet power, or some of the liberated radiation is advected along the jet or Compton scattered in the wind. 

Similarly extreme mass accretion rates as those seen in SS433 have been invoked in other objects, including TDEs (Kara et al. 2016, 2018), and rapidly growing high redshift quasars (Volonteri \& Rees 2005). Even where the accretion flow has low specific angular momentum (Bondi-like), disc accretion is still thought to take place (Inayoshi, Haiman \& Ostriker 2016) and may well resemble the super-critical discs we observe in ULXs and SS433 (see the discussion in Begelman \& Volonteri 2017 for the requisite conditions). However, neither of these classes of object are easy to study due to their inherently unpredictable appearance and rapidly diminishing brightness (TDEs), and distance (to both TDEs and quasars). We have demonstrated that in SS433 we can probe both the otherwise unseen super-critical inner regions and outflows through the application of established timing tools. Notably, in future, deep X-ray observations will allow us to model the phase-dependent lag in SS433, map the absorbing material in detail and thereby directly test theoretical predictions for super-critical accretion.

\section{Acknowledgements}

The authors thank the anonymous referee for their valuable suggestions. MJM and DJW appreciate support via STFC Ernest Rutherford Fellowships. WNA is supported by an ESA research fellowship. This research was partially supported by the Australian Government through the Australian Research Council's Discovery Projects funding scheme (project DP200102471). The authors thank Keith Arnaud for useful suggestions. This research has made use of data obtained with {\it NuSTAR}, a project led by Caltech, funded by NASA and managed by NASA/JPL, and has utilized the NUSTARDAS software package, jointly developed by the ASDC (Italy) and Caltech (USA). 

\section{Data Availability}

The data underlying this article were accessed from the HEASARC {\it NuSTAR} data repository at https://heasarc.gsfc.nasa.gov/docs/archive.html.  The derived data generated in this research will be shared on reasonable request to the corresponding author.

\bsp	% typesetting comment
\label{lastpage}


\vspace{-0.5cm}
\begin{thebibliography} {}

\bibitem[\protect\citeauthoryear{Abolmasov, Karpov \& Kotani}{2009}]{2009PASJ...61..213A} Abolmasov P., Karpov S., Kotani T., 2009, PASJ, 61, 213

\bibitem[\protect\citeauthoryear{Atapin, et al.}{2015}]{2015MNRAS.446..893A} Atapin K., Fabrika S., Medvedev A., Vinokurov A., 2015, MNRAS, 446, 893

\bibitem[\protect\citeauthoryear{Begelman, King \& Pringle}{2006}]{2006MNRAS.370..399B} Begelman M.~C., King A.~R., Pringle J.~E., 2006, MNRAS, 370, 399

\bibitem[\protect\citeauthoryear{Begelman \& Volonteri}{2017}]{2017MNRAS.464.1102B} Begelman M.~C., Volonteri M., 2017, MNRAS, 464, 1102

\bibitem[\protect\citeauthoryear{Blundell \& Bowler}{2005}]{2005ApJ...622L.129B} Blundell K.~M., Bowler M.~G., 2005, ApJL, 622, L129

\bibitem[\protect\citeauthoryear{Blundell, Bowler \& Schmidtobreick}{2008}]{2008ApJ...678L..47B} Blundell K.~M., Bowler M.~G., Schmidtobreick L., 2008, ApJL, 678, L47

\bibitem[\protect\citeauthoryear{Brinkmann, Kotani \& Kawai}{2005}]{2005A&A...431..575B} Brinkmann W., Kotani T., Kawai N., 2005, A\&A, 431, 575

\bibitem[\protect\citeauthoryear{Cherepashchuk et al.}{2020}]{2020NewAR..8901542C} Cherepashchuk A., Postnov K., Molkov S., Antokhina E., Belinski A., 2020, NewAR, 89, 101542. doi:10.1016/j.newar.2020.101542

\bibitem[\protect\citeauthoryear{Cherepashchuk, et al.}{2013}]{2013MNRAS.436.2004C} Cherepashchuk A.~M., Sunyaev R.~A., Molkov S.~V., Antokhina E.~A., Postnov K.~A., Bogomazov A.~I., 2013, MNRAS, 436, 2004

\bibitem[\protect\citeauthoryear{Cherepashchuk et al.}{2007}]{2007ESASP.622..319C} Cherepashchuk A.~M., Sunyaev K.~A., Seifina E.~V., Antokhina E.~A., Kosenko D.~I., Molkov S.~V., Shakura N.~I., et al., 2007, ESASP, 622, 319

\bibitem[\protect\citeauthoryear{Cherepashchuk, et al.}{2005}]{2005A&A...437..561C} Cherepashchuk A.~M., et al., 2005, A\&A, 437, 561


\bibitem[\protect\citeauthoryear{Dauser, Middleton, \& Wilms}{2017}]{2017MNRAS.466.2236D} Dauser T., Middleton M., Wilms J., 2017, MNRAS, 466, 2236 

\bibitem[\protect\citeauthoryear{Dolan, et al.}{1997}]{1997A&A...327..648D} Dolan J.~F., et al., 1997, A\&A, 327, 648

\bibitem[\protect\citeauthoryear{Eikenberry, et al.}{2001}]{2001ApJ...561.1027E} Eikenberry S.~S., Cameron P.~B., Fierce B.~W., Kull D.~M., Dror D.~H., Houck J.~R., Margon B., 2001, ApJ, 561, 1027

\bibitem[\protect\citeauthoryear{Fabian \& Rees}{1979}]{1979MNRAS.187P..13F} Fabian A.~C., Rees M.~J., 1979, MNRAS, 187, 13P

\bibitem[\protect\citeauthoryear{Fabrika, et al.}{2015}]{2015NatPh..11..551F} Fabrika S., Ueda Y., Vinokurov A., Sholukhova O., Shidatsu M., 2015, NatPh, 11, 551

\bibitem[\protect\citeauthoryear{Fabrika}{2004}]{2004ASPRv..12....1F} Fabrika S., 2004, ASPRv, 12, 1

\bibitem[\protect\citeauthoryear{Fabrika}{1997}]{1997Ap&SS.252..439F} Fabrika S.~N., 1997, Ap\&SS, 252, 439

\bibitem[\protect\citeauthoryear{Fuchs, Koch Miramond, \& {\'A}brah{\'a}m}{2006}]{2006A&A...445.1041F} Fuchs Y., Koch Miramond L., {\'A}brah{\'a}m P., 2006, A\&A, 445, 1041. doi:10.1051/0004-6361:20042160

\bibitem[\protect\citeauthoryear{Goranskij}{2011}]{2011PZ.....31....5G} Goranskij V., 2011, PZ, 31, 5

\bibitem[\protect\citeauthoryear{Harrison, et al.}{2013}]{2013ApJ...770..103H} Harrison F.~A., et al., 2013, ApJ, 770, 103

\bibitem[\protect\citeauthoryear{Inayoshi, Haiman \& Ostriker}{2016}]{2016MNRAS.459.3738I} Inayoshi K., Haiman Z., Ostriker J.~P., 2016, MNRAS, 459, 3738

\bibitem[\protect\citeauthoryear{Jiang, Stone, \& Davis}{2014}]{2014ApJ...796..106J} Jiang Y.-F., Stone J.~M., Davis S.~W., 2014, ApJ, 796, 106 

\bibitem[\protect\citeauthoryear{Jiang, Stone \& Davis}{2019}]{2019ApJ...880...67J} Jiang Y.-F., Stone J.~M., Davis S.~W., 2019, ApJ, 880, 67

\bibitem[\protect\citeauthoryear{Kaaret, Feng \& Roberts}{2017}]{2017ARA&A..55..303K} Kaaret P., Feng H., Roberts T.~P., 2017, ARA\&A, 55, 303

\bibitem[\protect\citeauthoryear{Kalberla, et al.}{2005}]{2005A&A...440..775K} Kalberla P.~M.~W., Burton W.~B., Hartmann D., Arnal E.~M., Bajaja E., Morras R., P{\"o}ppel W.~G.~L., 2005, A\&A, 440, 775

\bibitem[\protect\citeauthoryear{Kara, et al.}{2016}]{2016MNRAS.462..511K} Kara E., Alston W.~N., Fabian A.~C., Cackett E.~M., Uttley P., Reynolds C.~S., Zoghbi A., 2016, MNRAS, 462, 511

\bibitem[\protect\citeauthoryear{Kara, et al.}{2016}]{2016Natur.535..388K} Kara E., Miller J.~M., Reynolds C., Dai L., 2016, Natur, 535, 388

\bibitem[\protect\citeauthoryear{Kara, et al.}{2018}]{2018MNRAS.474.3593K} Kara E., Dai L., Reynolds C.~S., Kallman T., 2018, MNRAS, 474, 3593

\bibitem[\protect\citeauthoryear{Kara, et al.}{2020}]{2020MNRAS.491.5172K} Kara E., et al., 2020, MNRAS, 491, 5172

\bibitem[\protect\citeauthoryear{Kawai et al.}{1989}]{1989PASJ...41..491K} Kawai N., Matsuoka M., Pan H.-C., Stewart G.~C., 1989, PASJ, 41, 491

\bibitem[\protect\citeauthoryear{Khabibullin \& Sazonov}{2016}]{2016MNRAS.457.3963K} Khabibullin I., Sazonov S., 2016, MNRAS, 457, 3963

\bibitem[\protect\citeauthoryear{Khabibullin, Medvedev \& Sazonov}{2016}]{2016MNRAS.455.1414K} Khabibullin I., Medvedev P., Sazonov S., 2016, MNRAS, 455, 1414

\bibitem[\protect\citeauthoryear{King et al.}{2001}]{2001ApJ...552L.109K} King A.~R., Davies M.~B., Ward M.~J., Fabbiano G., Elvis M., 2001, ApJ, 552, L109 

\bibitem[\protect\citeauthoryear{King}{2009}]{2009MNRAS.393L..41K} King A.~R., 2009, MNRAS, 393, L41 

\bibitem[\protect\citeauthoryear{Kosec, et al.}{2018}]{2018MNRAS.479.3978K} Kosec P., et al., 2018, MNRAS, 479, 3978

\bibitem[\protect\citeauthoryear{Kotani, et al.}{1996}]{1996PASJ...48..619K} Kotani T., Kawai N., Matsuoka M., Brinkmann W., 1996, PASJ, 48, 619

\bibitem[\protect\citeauthoryear{Kotani, et al.}{1994}]{1994PASJ...46L.147K} Kotani T., et al., 1994, PASJ, 46, L147

\bibitem[\protect\citeauthoryear{Krivosheyev et al.}{2009}]{2009MNRAS.394.1674K} Krivosheyev Y.~M., Bisnovatyi-Kogan G.~S., Cherepashchuk A.~M., Postnov K.~A., 2009, MNRAS, 394, 1674. doi:10.1111/j.1365-2966.2009.14452.x

\bibitem[\protect\citeauthoryear{Kubota, et al.}{2010}]{2010PASJ...62..323K} Kubota K., et al., 2010, PASJ, 62, 323

\bibitem[\protect\citeauthoryear{Kubota et al.}{2007}]{2007ASPC..362..121K} Kubota K., Kawai N., Kotani T., Ueda Y., Brinkmann W., 2007, ASPC, 362, 121

\bibitem[\protect\citeauthoryear{Lockman, Blundell \& Goss}{2007}]{2007MNRAS.381..881L} Lockman F.~J., Blundell K.~M., Goss W.~M., 2007, MNRAS, 381, 881

\bibitem[\protect\citeauthoryear{Madsen, et al.}{2015}]{2015ApJS..220....8M} Madsen K.~K., et al., 2015, ApJS, 220, 8

\bibitem[\protect\citeauthoryear{Margon, et al.}{1979}]{1979ApJ...233L..63M} Margon B., Ford H.~C., Grandi S.~A., Stone R.~P.~S., 1979, ApJL, 233, L63

\bibitem[\protect\citeauthoryear{Margon \& Anderson}{1989}]{1989ApJ...347..448M} Margon B., Anderson S.~F., 1989, ApJ, 347, 448

\bibitem[\protect\citeauthoryear{Marshall, Canizares \& Schulz}{2002}]{2002ApJ...564..941M} Marshall H.~L., Canizares C.~R., Schulz N.~S., 2002, ApJ, 564, 941

\bibitem[\protect\citeauthoryear{Marshall, et al.}{2013}]{2013ApJ...775...75M} Marshall H.~L., et al., 2013, ApJ, 775, 75

\bibitem[\protect\citeauthoryear{Medvedev, et al.}{2018}]{2018AstL...44..390M} Medvedev P.~S., Khabibullin I.~I., Sazonov S.~Y., Churazov E.~M., Tsygankov S.~S., 2018, AstL, 44, 390

\bibitem[\protect\citeauthoryear{Medvedev \& Fabrika}{2010}]{2010MNRAS.402..479M} Medvedev A., Fabrika S., 2010, MNRAS, 402, 479

\bibitem[\protect\citeauthoryear{Medvedev, Khabibullin \& Sazonov}{2020}]{2020arXiv200512416M} Medvedev P., Khabibullin I., Sazonov S., 2020, arXiv, arXiv:2005.12416

\bibitem[\protect\citeauthoryear{Middleton et al.}{2011}]{2011MNRAS.411..644M} Middleton M.~J., Roberts T.~P., Done C., Jackson F.~E., 2011, MNRAS, 411, 644

\bibitem[\protect\citeauthoryear{Middleton et al.}{2014}]{2014MNRAS.438L..51M} Middleton M.~J., Walton D.~J., Roberts T.~P., Heil L., 2014, MNRAS, 438, L51

\bibitem[\protect\citeauthoryear{Middleton et al.}{2015}]{2015MNRAS.447.3243M} Middleton M.~J., Heil L., Pintore F., Walton D.~J., Roberts T.~P., 2015, MNRAS, 447, 3243

\bibitem[\protect\citeauthoryear{Middleton, et al.}{2015}]{2015MNRAS.454.3134M} Middleton M.~J., et al., 2015, MNRAS, 454, 3134

\bibitem[\protect\citeauthoryear{Middleton, et al.}{2018}]{2018MNRAS.475..154M} Middleton M.~J., et al., 2018, MNRAS, 475, 154

\bibitem[\protect\citeauthoryear{Middleton, et al.}{2019}]{2019MNRAS.489..282M} Middleton M.~J., Fragile P.~C., Ingram A., Roberts T.~P., 2019, MNRAS, 489, 282

\bibitem[\protect\citeauthoryear{Mu{\~n}oz-Darias, Motta \& Belloni}{2011}]{2011MNRAS.410..679M} Mu{\~n}oz-Darias T., Motta S., Belloni T.~M., 2011, MNRAS, 410, 679

\bibitem[\protect\citeauthoryear{Mushtukov, et al.}{2017}]{2017MNRAS.467.1202M} Mushtukov A.~A., Suleimanov V.~F., Tsygankov S.~S., Ingram A., 2017, MNRAS, 467, 1202

\bibitem[\protect\citeauthoryear{Narayan, S{\c a}dowski, \& Soria}{2017}]{2017MNRAS.469.2997N} Narayan R., S{\c a}dowski A., Soria R., 2017, MNRAS, 469, 2997 

\bibitem[\protect\citeauthoryear{Ohsuga \& Mineshige}{2011}]{2011ApJ...736....2O} Ohsuga K., Mineshige S., 2011, ApJ, 736, 2

\bibitem[\protect\citeauthoryear{Pinto, Middleton, \& Fabian}{2016}]{2016Natur.533...64P} Pinto C., Middleton M.~J., Fabian A.~C., 2016, Natur, 533, 64 

\bibitem[\protect\citeauthoryear{Pinto et al.}{2017}]{2017MNRAS.468.2865P} Pinto C., et al., 2017, MNRAS, 468, 2865 

\bibitem[\protect\citeauthoryear{Pinto, et al.}{2020}]{2020MNRAS.491.5702P} Pinto C., et al., 2020, MNRAS, 491, 5702

\bibitem[\protect\citeauthoryear{Pintore, et al.}{2017}]{2017ApJ...836..113P} Pintore F., Zampieri L., Stella L., Wolter A., Mereghetti S., Israel G.~L., 2017, ApJ, 836, 113

\bibitem[\protect\citeauthoryear{Poutanen et al.}{2007}]{2007MNRAS.377.1187P} Poutanen J., Lipunova G., Fabrika S., Butkevich A.~G., Abolmasov P., 2007, MNRAS, 377, 1187 

\bibitem[\protect\citeauthoryear{Rees}{1988}]{1988Natur.333..523R} Rees M.~J., 1988, Natur, 333, 523

\bibitem[\protect\citeauthoryear{Revnivtsev, et al.}{2004}]{2004A&A...424L...5R} Revnivtsev M., et al., 2004, A\&A, 424, L5

\bibitem[\protect\citeauthoryear{Revnivtsev, et al.}{2006}]{2006A&A...447..545R} Revnivtsev M., et al., 2006, A\&A, 447, 545

\bibitem[\protect\citeauthoryear{Revnivtsev, Gilfanov \& Churazov}{1999}]{1999A&A...347L..23R} Revnivtsev M., Gilfanov M., Churazov E., 1999, A\&A, 347, L23

\bibitem[\protect\citeauthoryear{S{\c a}dowski et al.}{2014}]{2014MNRAS.439..503S} S{\c a}dowski A., Narayan R., McKinney J.~C., Tchekhovskoy A., 2014, MNRAS, 439, 503 

\bibitem[\protect\citeauthoryear{Shakura \& Sunyaev}{1973}]{1973A&A....24..337S} Shakura N.~I., Sunyaev R.~A., 1973, A\&A, 24, 337

\bibitem[\protect\citeauthoryear{Shklovsky}{1981}]{1981AZh....58..554S} Shklovsky I.~S., 1981, AZh, 58, 554

\bibitem[\protect\citeauthoryear{Silva, Uttley, \& Costantini}{2016}]{2016A&A...596A..79S} Silva C.~V., Uttley P., Costantini E., 2016, A\&A, 596, A79 

\bibitem[\protect\citeauthoryear{Singh, White \& Drake}{1996}]{1996ApJ...456..766S} Singh K.~P., White N.~E., Drake S.~A., 1996, ApJ, 456, 766

\bibitem[\protect\citeauthoryear{Soria, et al.}{2014}]{2014Sci...343.1330S} Soria R., et al., 2014, Sci, 343, 1330

\bibitem[\protect\citeauthoryear{Soria, et al.}{2020}]{2020ApJ...888..103S} Soria R., Blair W.~P., Long K.~S., Russell T.~D., Winkler P.~F., 2020, ApJ, 888, 103

\bibitem[\protect\citeauthoryear{Stephenson \& Sanduleak}{1977}]{1977ApJS...33..459S} Stephenson C.~B., Sanduleak N., 1977, ApJS, 33, 459

\bibitem[\protect\citeauthoryear{Sutton, Roberts \& Middleton}{2013}]{2013MNRAS.435.1758S} Sutton A.~D., Roberts T.~P., Middleton M.~J., 2013, MNRAS, 435, 1758

\bibitem[\protect\citeauthoryear{Takeuchi, Ohsuga, \& Mineshige}{2013}]{2013PASJ...65...88T} Takeuchi S., Ohsuga K., Mineshige S., 2013, PASJ, 65, 88 

\bibitem[\protect\citeauthoryear{Urquhart \& Soria}{2016}]{2016ApJ...831...56U} Urquhart R., Soria R., 2016, ApJ, 831, 56

\bibitem[\protect\citeauthoryear{Uttley et al.}{2014}]{2014A&ARv..22...72U} Uttley P., Cackett E.~M., Fabian A.~C., Kara E., Wilkins D.~R., 2014, A\&ARv, 22, 72 

\bibitem[\protect\citeauthoryear{van den Heuvel}{1981}]{1981VA.....25...95V} van den Heuvel E.~P.~J., 1981, VA, 25, 95

\bibitem[\protect\citeauthoryear{van Velzen \& Farrar}{2014}]{2014ApJ...792...53V} van Velzen S., Farrar G.~R., 2014, ApJ, 792, 53

\bibitem[\protect\citeauthoryear{Vaughan \& Nowak}{1997}]{1997ApJ...474L..43V} Vaughan B.~A., Nowak M.~A., 1997, ApJL, 474, L43

\bibitem[\protect\citeauthoryear{Vermeulen, et al.}{1987}]{1987Natur.328..309V} Vermeulen R.~C., Schilizzi R.~T., Icke V., Fejes I., Spencer R.~E., 1987, Natur, 328, 309

\bibitem[\protect\citeauthoryear{Verner, Verner \& Ferland}{1996}]{1996ADNDT..64....1V} Verner D.~A., Verner E.~M., Ferland G.~J., 1996, ADNDT, 64, 1

\bibitem[\protect\citeauthoryear{Volonteri \& Rees}{2005}]{2005ApJ...633..624V} Volonteri M., Rees M.~J., 2005, ApJ, 633, 624

\bibitem[\protect\citeauthoryear{Waisberg, et al.}{2019}]{2019A&A...624A.127W} Waisberg I., Dexter J., Olivier-Petrucci P., Dubus G., Perraut K., 2019, A\&A, 624, A127

\bibitem[\protect\citeauthoryear{Walton, et al.}{2020}]{2020MNRAS.tmp.1280W} Walton D.~J., et al., 2020, MNRAS.tmp, doi:10.1093/mnras/staa1129

\bibitem[\protect\citeauthoryear{Walton, et al.}{2018}]{2018ApJ...856..128W} Walton D.~J., et al., 2018, ApJ, 856, 128

\bibitem[\protect\citeauthoryear{Walton et al.}{2016}]{2016ApJ...826L..26W} Walton D.~J., et al., 2016, ApJ, 826, L26

\bibitem[\protect\citeauthoryear{Wilkinson \& Uttley}{2009}]{2009MNRAS.397..666W} Wilkinson T., Uttley P., 2009, MNRAS, 397, 666

\bibitem[\protect\citeauthoryear{Yuan et al.}{1995}]{1995A&A...297..451Y} Yuan W., Kawai N., Brinkmann W., Matsuoka M., 1995, A\&A, 297, 451


\end{thebibliography}
\end{document}